\documentclass[12pt]{article}
\usepackage{geometry}
\usepackage[utf8]{inputenc}

 \usepackage{amssymb,amsmath,epsfig}

\begin{document}

\title{\textbf{Relativistic Polytropic Models of Charged Anisotropic Compact Object}}

\author{H. Nazar$^{a}$ \thanks{hammadnazar350@gmail.com}, M. Azam $^{b}$\thanks{azam.math@ue.edu.pk, azammath@gmail.com}, G. Abbas$^{a}$ \thanks{ghulamabbas@iub.edu.pk}, Riaz Ahmed$^c$ \thanks{rahmed@miu.edu} and R. Naeem$^b$ \thanks{go4rabia@gmail.com}\\
$^a$Department of Mathematics The Islamia University\\ of Bahawalpur,
Bahawalpur, Pakistan
\\$^b$ Department of Mathematics,\\ Division of Science and Technology,\\
University of Education, Lahore, Pakistan
\\$^c$ Department
 of Mathematics,\\ Maharishi International University,  Fairfield, IA USA.}

\date{}

\maketitle
\begin{abstract}
In this paper, we have introduced new viable solutions of Einstein-Maxwell field equations by incorporating the features of anisotropic matter distribution in the realm of General theory of Relativity ($GR$). For this procurement, we have employed a Finch-Skea spacetime along with a generalized polytropic equation of state ($EoS$). We have constructed various models of generalized polytropes by assuming the different choices of the polytropic index i.e.,$\eta=\frac{1}{2}, \frac{2}{3}, 1$ and $2$. The numerous physical characteristics of these considered models have been studied via graphical analysis, which obey all the essential conditions of the astrophysical compact objects. Furthermore, such outcomes of charged anisotropic compact star models can be regained to the various cases such as linear, quadratic and polytropic $EoS$.
\end{abstract}
{\bf Keywords:} General Relativity, Exact solutions, Charged polytropic models, Anisotropy of pressure;
{\section{Introduction}
By using the basic formulation of $GR$, the investigation of stellar astrophysical compact objects ($COs$) is still a growing and interesting field of research. It is essential for the modeling of these gravitational $COs$ to be static or non-static and spherically symmetric regarding to observe their stable configuration. To develop a realistic and potentially stable model of astrophysical $CO$ together with the distribution of an electric charge is ever being a solid choice among the theorists and astrophysicists. It is necessary to tackle the system of Einstein-Maxwell field equations to reduce into a less complex system when someone adopts a better mathematical tool. However, numerous researchers have chosen distinctive ansatz to comprehend these equations.
In this view point, many authors such as Patel and Kopper \cite{1}, Patel et al. \cite{2}, Tikekar and Singh \cite{3}, Sharma et al. \cite{4} and Komathiraj and Maharaj \cite{5} have used a variety of numerical strategies to get the definite solutions for the relativistic $COs$. These investigations reflect that the system of field equations is necessarily depicted as astrophysical $COs$.

In $GR$, the exact solutions are physically viable and robustly utilized in numerous applications of astrophysical $COs$. To attain the analytical solutions of the Einstein-Maxwell equations is one of the basic problems in the $GR$ as well as among the modified versions of this theory of gravitation. It is crucial to guarantee that the physical nature of these solutions should be valid. In this perspective, the physical effectiveness of exact solutions has been manifested by Delgaty and Lake \cite{6}. They have fixed the geometrical standards for the gravitational potential, which imperatively fulfill all the physical traits.

The presence of electric charge within the matter content creates an extraordinary theoretical interest for the electromagnetic effects on the stellar astrophysical $COs$. Gupta and Maurya \cite{7, 9} have developed the super-dense star models in the presence of electric intensity. They have analyzed that the obtained solutions are well-behaved and regular within the entire domain of the star. Maurya and Gupta \cite{10} have proposed some vital features of charged objects by assigning the value of electric parameter. Pant et al. \cite{11} have explored various solutions of charge matter balls influence with finite pressure and matter density that satisfied the casuality condition $(\frac{d P_{r}}{d \rho}\leq 1)$ and also possessed the positive nature at the center. Some well-behaved charged $COs$ have been investigated by Maurya and Gupta \cite{12, 13} with a metric potential $g_{44}=B (1+ C r^{2})^{n}$. Beside of these exceptional work, several researchers \cite{12}-\cite{15} have contributed a variety of novel solutions regarding the modeling of charged $COs$.

During the recent decades, numerous authors have given the exact solutions of Einstein-Maxwell field equations with isotropic fluid. Some global and physically realistic solutions of charged isotropic and Reissner-Nordstr$\ddot{o}$m metric were studied by Ivanov \cite{16}. In order to obtain the exact solution, one needs to integrate the Einstein-Maxwell field equations. Therefore, distinctive confinement has been set by researchers on the spacetime geometry as well as on the matter distribution. Under the supposition of constant hypersurface, Komathiraj and Maharaj \cite{17} have introduced the charged spherical models. By setting the constraint on gravitational potential, Thirukkanesh and Maharaj \cite{18} and Komathiraj and Maharaj \cite{5} have investigated the exact solutions of the Einstein-Maxwell field equations in the presence of isotropic matter. Mak and Harko \cite{19} have proposed the different techniques to obtain the various physical aspects of the isotropic spheres. In contrast to these inspirational consequences, many researchers have shown that the pressure anisotropy is an underlying ingredient in the modeling of $COs$. The modeling of anisotropic $COs$ is still an extensive and large dynamical field among the research groups during the last several decades. In this sense, Ruderman \cite{20} provided the basic foundation for the configuration of stellar anisotropic $COs$ and concluded that the extreme dense regions of $COs$ contain anisotropic fluid. Further, Bowers and Liang \cite{21} have studied the possible causes of anisotropic stress. Subsequently, the impact of anisotropic pressure on the system of general relativistic sphere and its physical characteristics have been investigated by Dev and Gleiser \cite{22, 23}. Herrera and his co-researchers \cite{23s1}-\cite{24} have considered a class of exact solutions to the set of field equations regarding the physical emergence of anisotropic pressure in the self-gravitating structures. Moreover, Fuloria and Pant \cite{26} and Maurya and Maharaj \cite{27} have suggested the various types of analytical solutions to demonstrate the physical implications of anisotropic compact spheres in the regime of Karmarkar approach. Apart of these notable consequences, some class of exact relativistic solutions to the Einstein field equations for anisotropic spherically symmetric and static dense compact star models inspired by the Buchdahl and Korkina-Orlyanskii ansatzes have been proposed in \cite{27a}-\cite{27d}. Recently, many researcher have made phenomenological considerations in the assessment of physical influences of anisotropic stress in the realm of $GR$ as well as alternative gravitational theories \cite{s9}-\cite{s27}.

No doubt, it is a basic issue to choose a suitable $EoS$ for the demonstration of astrophysical $COs$. For instance, Varela et al. \cite{28} have discussed the physical reliability of $EoS$ in the demonstration of $COs$. Sharma and Maharaj \cite{29}, Thirukkanesh and Maharaj \cite{30} and Takisa and Maharaj \cite{31} have employed the linear $EoS$ to develop the models of dense $COs$. For spherically symmetric systems, Hansraj and Maharaj \cite{32} have studied the isotropic matter distribution that satisfied the barotropic $EoS$, i.e., $ p_{r}=p_{r}(\rho)$. In context of quadratic $EoS$, Feroze and Siddique \cite{33} and Maharaj and Takisa \cite{34} have explored the physical implications of charged anisotropic relativistic star models. In the composition of astronomical compact star models, the polytropic $EoS$ plays a vital role to explain the various features of compact stars. Chandrasekhar \cite{35} have discussed the outcomes of Newtonian polytropic theory in the context of thermodynamical laws. Herrera and Barreto \cite{36} have organized a detail study for the Newtonian polytropic models when the fluid is considered as anisotropic. Thirukkanesh and Ragel \cite{37,38} have depicted the realistic features of uncharged compact star models in the regime of polytropic $EoS$. Furthermore, Takisa and Maharaj \cite{39} have reviewed the physical consequences of polytropic $EoS$ on charged anisotropic compact star models. Nilsson and Ugla \cite{40} have adopted the standard procedure to obtain a set of equations by admitting the polytropic $EoS$. This $EoS$ can be assessed numerically to finite radius corresponding to the various values of polytropic index $\eta$ defined from $0$ to 3.339 and for $\eta\geq5$ to infinite radius.

Despite of the above chosen polytropic $EoS$, the combination of linear and polytropic $EoS$ gives the generalized form of polytropic $EoS$ i.e., $ P_{r}=\alpha \rho + \beta \rho^{1+1/\eta}$. Several authors have employed the generalized polytropic $EoS$ to study the definite stellar features of $COs$ like as, Chavanis \cite{41} constructed the models of the early and the late Universe with a generalized polytropic $EoS$ and concluded that the models can be seem in fine symmetric tune for an early and late epoch of inflation. Azam et al. \cite{43} have formulated the general framework to explore the behavior and reliability of Newtonian and relativistic polytropes in the presence of generalized polytropic $EoS$. Nasim and Azam \cite{45} have proposed the physically viable charged anisotropic models by considering the generalized polytropic $EoS$. To model the ultra relativistic $CO$, Finch and Skea \cite {46} have developed the analytical study in the background of Duorah and Ray ansatz. Considering the Finch-Skea ansatz, some viable implications for charged anisotropic compact star models have been proposed by Hansraj and Maharaj \cite{32} by employing the barotropic $EoS$. Sharma and Ratanpal \cite{47} have reviewed the various fruitful outcomes of static and spherically symmetric objects in the framework of Finch-Skea ansatz. Pandya et al. \cite{48} have shown the generalized results of Finch-Skea geometry for the modeling of strange quark stars by considering the Chaplygin gas $EoS$. Recently, several notable characteristics of astrophysical $COs$ have been obtained by assuming the Finch-skea geometries in various gravitational theories \cite{48a}-\cite{48e}.

This manuscript is outlined as follows: In Sec. \textbf{2}, we have investigated an equivalent set of Einstein-Maxwell field equations for models of anisotropic astrophysical $COs$ by implementing a generalized polytropic $EoS$  with the transformation introduced by Durgapal and Bannerji \cite{49}. The smooth matching conditions have also been investigated between the interior metric and exterior Reissner-Nordstr$\ddot{\ddot{o}}$m solution at the hypersurface. In Sec. \textbf{3}, we have assumed a gravitational potential $Z$ as Finch-Skea ansatz, which leads us to a restricted system of field equations in comprehensive manner. We have obtained the various feasible class of charged polytropic models corresponding to the particular choices of polytropic index $\eta$ in Sec. \textbf{4}. In Sec. \textbf{5}, deals with physical adequacy of presented models to study the various attributes such as, material quantities, effective mass function, anisotropy, electric charge and causality condition. The last section includes the summary of our findings.
\section{Maxwell field equations}
We choose a static and spherically symmetric relativistic stellar interior model in the standard coordinates as $x^{a}=(x^{1}, x^{2}, x^{3}, x^{4})=(t, r, \theta, \phi)$, which has the form
\begin{equation}\label{1}
ds^2=-e^{2\lambda(r)}dt^2+e^{2\psi(r)}dr^{2}+r^{2}d\Omega^2,
\end{equation}
where $\lambda(r)$ and $\psi(r)$ are metric potentials for a static gravitational field, and $d\Omega^{2}=d\theta^{2}+\sin\theta^{2} d\phi^2$. We assume that the interior region of the stellar model is filled by the charged anisotropic matter, whose total energy-momentum tensor is given by
\begin{equation}\label{2}
T_{a}^{b}=diag \Big[-\rho-\frac{\mathcal{E}^{2}}{2}, P_{r}-\frac{\mathcal{E}^{2}}{2}, P_{t}+\frac{\mathcal{E}^{2}}{2}, P_{t}+\frac{\mathcal{E}^{2}}{2}\Big].
\end{equation}
Here $\rho$ illustrates the energy density, $P_{r}$ and $P_{t}$ are radial and tangential pressure, respectively, whereas $\mathcal{E}$ is the electric field intensity. For static interior compact sphere (\ref{1}) and anisotropic energy-momentum tensor (\ref{2}), the Einstein-Maxwell field equations get the following form
\begin{eqnarray}\label{3}
8 \pi \ \rho+\mathcal{E}^{2}&=&\frac{1}{r^2}\ \Big[r(1-e^{-2\psi})\Big]' \label{4},\\
8 \pi \ P_{r}-\mathcal{E}^{2}&=&-\frac{1}{r^2}\ \Big(1- e^{-2\psi}\Big)+\frac{2\lambda'}{r}e^{-2\psi} \label{5},\\
8 \pi \ P_{t}+ \mathcal{E}^{2}&=&e^{-2\psi}\ \Big(\lambda''+\lambda'^{2}+\frac{\lambda'}{r}-\lambda' \psi'-\frac{\psi'}{r}\Big),\\ \label{6}
\sigma&=&\frac{1}{r^2}\ e^{-\psi}\ \Big(r^2\ \mathcal{E}\Big)'.
\end{eqnarray}
Here, we use the geometric units as $G = c = 1$, whereas $'$ and $''$ signs are first and second derivatives with respect to $r$. The underlying equations describing the fundamental stellar interior model for charged anisotropic fluid sphere are provided by the system of Eqs. (\ref{3})-(\ref{6}). These Einstein-Maxwell field equations are a system of four independent equations in six independent unknowns ($\lambda$, $\psi$, $\rho$, $P_{r}$, $P_{t}$, $\mathcal{E}$ or $\sigma$). Remember, if we assume a specific form of the electric field intensity $\mathcal{E}$ then the system of Eqs. (\ref{3})-(\ref{6}) becomes a system of three independent equations in five independent unknowns. A charged solution can be generated by specifying the forms for two unknown functions  or each combination of unknown functions and equations of state associating the matter quantities. In order to evolve a simplest system, we consider an important relation between the energy density and radial pressure, which is described by the generalized polytropic $EoS$.
\begin{eqnarray}\label{7}
P_{r}=\mathcal{A}\ \rho+\mathcal{B}\ \rho^{\Gamma},
\end{eqnarray}
where $\Gamma= 1+\frac{1}{\eta}$ in which $\eta$ refers to the polytropic index, while $\mathcal{A}$ and $\mathcal{B}$ are arbitrary constants. In deciding criterions for a certain $EoS$ to construct the stellar relativistic astrophysical $CO$ rather than the self-gravitating compact model the radial pressure $P_{r}$ should be non-negative and regular (finite) at every point inside the stars and it should be vanished at the surface $r=r_{b}$ of the sphere \cite{49a}. Therefore, when pressure is zero at the boundary, implies that the density also vanishes on the surface of the star. To obtain a concise system of the field equations, we imply the transformation priory proposed by Durgapal and Banerji \cite{49} given in the following way
\begin{equation}\label{8}
x=r^{2},\ \ \      \mathcal{Z}(x)= e^{-2\psi(r)},\ \ \       y^{2}(x)= e^{2\lambda(r)},
\end{equation}
Using the above transformation, Eqs(\ref{3})-(\ref{6}) along with the generalized polytropic $EoS$ can be rewritten as
\begin{eqnarray}\label{9}
8 \pi\rho&=&\frac{1}{x}(1-\mathcal{Z})-2\dot{\mathcal{Z}}-\mathcal{E}^{2},\\
P_{t}&=&P_{r}+\Delta\label{10},\\
P_{r}&=&\mathcal{A}\ \rho + \mathcal{B}\ \rho^{1+\frac{1}{\eta}}\label{11},\\
8 \pi P_{r}&=&4\mathcal{Z}\frac{\dot{y}}{y}+\frac{1}{x}\Big(\mathcal{Z}-1\Big)+\mathcal{E}^{2}\label{12},\\
8 \pi P_{t}&=&4x\mathcal{Z}\frac{\ddot{y}}{y} +\Big(4\mathcal{Z}+2x\dot{\mathcal{Z}}\Big)\frac{\dot{y}}{y}+\dot{\mathcal{Z}}-\mathcal{E}^{2} \label{13},\\
8\pi\Delta&=&4x\mathcal{Z}\frac{\ddot{y}}{y}+2x\frac{\dot{y}}{y}\dot{\mathcal{Z}}+\dot{\mathcal{Z}}+\frac{1}{x}\Big(1-\mathcal{Z}\Big)-2\mathcal{E}^{2} \label{14},\\
\sigma^{2}&=&\frac{4\mathcal{Z}}{x}\Big(\mathcal{E}+\dot{\mathcal{E}}x\Big)^{2}\label{15}.
\end{eqnarray}
From Eqs.(\ref{7}-(\ref{9}), we get
\begin{eqnarray}\label{16}
\frac{\dot{y}}{y}&=&\frac{\mathcal{A}}{4\mathcal{Z}}\Big(\frac{1-\mathcal{Z}}{x}-2\dot{\mathcal{Z}}-\mathcal{E}^{2}\Big)+\frac{\mathcal{B}(8\pi)^{-\frac{1}{\eta}}}
{4\mathcal{Z}}\Big(\frac{1-\mathcal{Z}}{x}-2\dot{\mathcal{Z}}-\mathcal{E}^{2}\Big)^{1+\frac{1}{\eta}}+\frac{1-\mathcal{Z}}{4\mathcal{Z}x}-\frac{\mathcal{E}^{2}}{4\mathcal{Z}},
\end{eqnarray}
where dot represents differentiation with respect to $x$, and anisotropy factor is $\Delta$, which reflect the difference between the
radial and the tangential pressure. In this description, Thirukkanesh and Maharaj \cite{30} have assumed the linear $EoS$ to develop an analogous equatorial system for the modeling of anisotropic dense compact sphere. Feroze and Siddique \cite{33} and Maharaj and Takisa \cite{34} have employed the quadratic $EoS$ to obtain the physically realistic compact star models. Furthermore, the polytropic $EoS$ is assumed by Takisa and Maharaj \cite{31} in order to construct a charged anisotropic compact star model. The gravitating mass within the sphere of radius $r$ is expressed as
\begin{equation}\label{17}
M(r)=4\pi \int^r_0 \omega^2 \rho(\omega) d\omega.
\end{equation}
The mass function after implementing the transformation takes the following form
\begin{equation}\label{18}
M(x)=2\pi\int^x_0 \sqrt{\omega}\rho(\omega)d\omega.
\end{equation}
\subsection{Matching conditions}
The matching conditions are generally employed to join interior and exterior geometries of the star at the boundary surface $r=r_{b}$. The choice of the exterior region entirely depends on the interior matter distribution of the star. For spheroidal structure, if the fluid in the interior of an object is electrically charged and anisotropic, we assume the Reissner-Nordstr$\ddot{\ddot{o}}$m metric as the exterior region,
\begin{equation}\label{18a}
ds^2=-\Big(1-\frac{2\tilde{M}}{r_{b}}+\frac{Q^2}{r_{b}^{2}}\Big)dt^2+\Big(1-\frac{2\tilde{M}}{r_{b}}+\frac{Q^2}{r_{b}^{2}}\Big)^{-1}dr^2
+r_{b}^{2}\Big(d\theta^2+\sin^2\theta d\phi^2\Big),
\end{equation}
where $\tilde{M}$ is the total mass of the stellar interior and $Q$ indicates the total electric charge of the fluid. Matching conditions play a vital role in the study of the compact star. These conditions tell whether the junction of two geometries give a realistic solution or not when a boundary surface separates the region into inner and outer regions. The smooth matching of inner and outer metrics via the continuity of the first and second fundamental form over the boundary surface \cite{49b}-\cite{49d} determines the following results:
\begin{eqnarray}\nonumber
e^{2\lambda(r)}&=&e^{-2\psi(r)}=\Big(1-\frac{2\tilde{M}}{r_{b}}+\frac{Q^2}{r_{b}^{2}}\Big),~~~~~\textit{M}=\tilde{M},~~~~~q=Q,\\
P_{r}&=&0.\label{18b}
\end{eqnarray}
The mass function of the stellar interior $CO$ defined by Misner and Sharp \cite{49a} and Nielsen and Yeom \cite{49e}, is given by
\begin{eqnarray}\label{18c}
\textit{M}&=&\frac{r}{2}\Big(1-\frac{1}{e^{2\psi}}+\frac{q^{2}}{r^{2}}\Big),
\end{eqnarray}
where $q=2\pi\int_{0}^{x}\sigma e^{\psi(x)}\sqrt{x}dx$ is the total charge enclosed by a sphere of radius $r$. In spheroidal system as a bounded object the mass of the star can be computed as a measure of the total energy within a sphere of a radius $r$. The concept of polytropes is based on the presumption of hydrostatic stability and polytropic $EoS$. In this analysis, we study the polytropic models
in the scenario of a generalized polytropic $EoS$, which is a combination of linear and polytropic $EoS$.
\section{Comprehensive overview on Finch-Skea ansatz}
In this perspective, our fundamental endeavor is to develop a fine tune model for stable configuration of the stellar relativistic dense object, when no expedient proofs concerning the evolution and nature of the particle interactions are commodious. It might be formulated by determining the viable solutions of equations of motion describing the static stellar interior core of spherical $CO$. However, analytical solution of equations of motion is not an easy task because of the highly non-linear differential equations. In this view point, several appropriate approaches have been successfully imposed to tackle the system of differential equations. Since, Einstein's field equations permits a joint affinement between the matter field and its geometry, we shall affirm a structural approach to deal with such a constraint. To do so, an authentic form of one of the metric potentials with clear attribution of an analogous metric will be specified to point out the other. Such approach was developed by Finch and Skea \cite{46} for the composition of an interior spheroidal geometry. Duorah and Ray \cite{50} were the pioneer, who gave the idea of such ansatz in which they did not provide an ideal form of such ansatz to meet the equations of motion for the modeling of astrophysical $COs$. This type of ansatz has procured lot of insights in the composition of astrophysical $COs$ and consequences are much suitable and satisfy all the constraints, which are necessary for the acceptability of the models \cite{6}. Various fruitful findings regarding to the such Finch-Skea ansatz have been reviewed to identify an extensive group of paradigms for the stellar astrophysical $COs$ by incorporating electric charge, dissipative matter content, pressure anisotropy, quadratic $EoS$, quintessence matter, etc. \cite{32,47,48,50a,50b}.

Inspiring from this motivational information, we reported here that in the connection of such ansatz too many realistic features of cosmological and stellar astrophysical $COs$ have been studied in $GR$ as well as in alternative theories of gravitation. In the discipline of the stellar astrophysical bodies, two parameters viable class of an exact solutions of compact relativistic star model in the presence of strange quark source have proposed by Tikekar and Jotania \cite{50c}. Bhar \cite{50d} has presented the implications of spherical $CO$ supported by the Chaplygin $EoS$ and examined that the $CO$ is composed of a quark matter content. Banerjee et al. \cite{50e} have contemplated the viable features of an astrophysical compact body with an exterior $BTZ$ metric under the framework of Finch-Skea ansatz. Bhar et al. \cite{50f} have explored a viable class of analytical solutions of Einstein field equation by using the $MIT$ bag $EoS$ in the presence of Finch-Skea geometry for the formulation of stellar relativistic compact sphere. In additionally, various theoretical physicists have done a variety of novel work by applying the Finch-Skea ansatz as well as its generalization in the higher dimensional gravity theories \cite{50g}-\cite{50k}. In spite of the astrophysical implications, a quite fascinating and qualitative analysis has been disclosed in the regime of Finch-Skea geometries regarding to observe the cosmological evidences by adopting the Chaplygin gas model, barotropic, quadratic and MIT bag $EoS$. These explicit models of $EoS$ have described the immense influence of matter content in the forms of dark matter and dark energy, because they have robustly affirmed our thoughts regarding the conceptions of unseen features of inflationary Universe. Over the last few decades, some researchers have reviewed the models of Chaplygin gas $EoS$ into the account of evolutionary expansion of the Universe and composition of massive scale objects \cite{50l}-\cite{50m}. In some of recent investigations, Chanda et al. \cite{48d} have proposed quite fascinating work within the $f(T)$ theory of gravitation for the modeling of stellar anisotropic $COs$ by implementing the Finch-Skea geometry. Singh et al. \cite{48e} have proposed a class of analytical solutions to formulate a model of anisotropic compact sphere by using the Finch-Skea ansatz. Maurya and his collaborators \cite{50n}-\cite{50q} have investigated various viable class of an exact solutions for the modeling of charged and without charged anisotropic relativistic $COs$ by proposing the MIT bag model $EoS$ and Finch-Skea ansatz within the context of $GR$ and alternative theories of gravity. They explored that under described conditions these solutions are well behaved and satisfy all the necessary bounds for models of stellar anisotropic $CO$. After an inspirational overview on such ansatz, now our basic purpose to develop a new family of an exact solutions for the Einstein-Maxwell field equations. For this purpose, we need to assume one of the metric potentials as $\mathcal{Z}$ given in Ref.\cite{46}. Hence, we take
\begin{equation}\label{19}
e^{2\psi}= 1+\frac{r^2}{R^2},
\end{equation}
where $R$ describes the curvature parameter. The geometric approach of such solution has been successfully imposed to satisfy the necessary conditions for the composition of stellar relativistic $COs$. By using Eq. (\ref{8}) and (\ref{19}), we get
\begin{equation}\label{20}
\mathcal{Z}(x)=e^{-2\psi}= \frac{1}{1+\frac{r^2}{\textit{R}^2}}.
\end{equation}
We assume a real arbitrary constant $\xi$ in above form, which gives
\begin{equation}\label{21}
\mathcal{Z}(x)=({1+\xi x})^{-1},\  \  \  \   \xi \neq0.
\end{equation}
The ansatz $\mathcal{Z}$ is physically admissible as it is non-singular and continuous at the central core of star. Maharaj et al \cite{51} have studied new feasible class of analytical solutions in the Finch-Skea spacetime to develop a model of anisotropic fluid sphere. Sharma et al. \cite{52} have constructed  singularity free solutions for anisotropic compact body by setting a parameter $\xi=1$ within the Finch-Skea spacetime. Consequently, the choice of ansatz $\mathcal{Z}$ is trustworthy to obtain physically realistic charged sphere models having anisotropic fluid configuration through generalized polytropic $EoS$. Our consideration for electromagnetic charge $\mathcal{E}$ is
\begin{equation}\label{22}
\mathcal{E}^{2}=\frac{\alpha x}{(1+\xi x)^{2}}.
\end{equation}
Here $\alpha $ is a real arbitrary constant. For modeling of charged relativistic $CO$, it is vital to make sure that two generic aspects are contained in the model. First, the model must be physically acceptable, i.e., the metric potentials, electric charge and matter components should be regular and free from physical singularities within the entire distribution of the sphere, the inner region joins smoothly with the outer metric and causality is not breached. Finally, we must retrieve neutral solution (uncharged) of the equations of motion when the electric charge vanishes, an uncharged sphere must be retrievable as a potentially stable final fate. Due to the closed bounded interior configuration, the above choice of $\mathcal{E}$ is quite physically reliable and sustainable to predict a potentially viable astrophysical compact star model. For the appropriate choice of independent variable $x$, the choice of $\mathcal{E}$ gives the constant term. Hansraj and Maharaj \cite{32} and Nasim and Azam \cite{45} have applied analogous choice of an electric charge to compose a potentially stable charged anisotropic compact star models. Moreover, Maurya et al. \cite{52v} have investigated the singularity free solution of charged anisotropic fluid sphere by assuming the specific choice of electric intensity into an account of $GR$. Consequently, this choice is most capable to construct these models with generalized polytropic $EoS$ in spacetime geometry of Finch and Skea. Hence, Eqs.(\ref{9})-(\ref{16}) become
\begin{eqnarray}\label{23}
\rho &=& \frac{3\xi+\xi^2 x-\alpha x}{8 \pi(1+\xi x)^{2}}\label{24}, \\
P_{r} &=& \frac{\mathcal{A}}{8 \pi}\Big[\frac{3\xi+\xi^2 x-\alpha x}{(1+\xi x)^{2}}\Big]+\frac{\mathcal{B}}{(8\pi)^{1+ \frac{1}{\eta}}} \Big[\frac{3\xi+\xi^2 x- \alpha x}{(1+\xi x)^{2}}\Big]^{1+\frac{1}{\eta}}\label{25}, \\
\sigma^{2}&=&\frac{\alpha\ (3+\xi x)^{2}}{(1+\xi x)^{5}}\label{26}, \\
P_{t} &=& P_{r}+\Delta,
\end{eqnarray}
whereas, the pressure anisotropy for our charged interior compact sphere model can be described as
\begin{eqnarray}\label{27}
8\pi\Delta&=&\frac{4 x}{1+\xi x}\Big(\frac{\dot{y}}{y}\Big)^{2}-\frac{2\xi x}{(1+\xi x)^{2}}\Big(\frac{\dot{y}}{y}\Big)+\frac{4 x}{1+\xi x}\Big[\frac{d}{dx}\Big(\frac{\dot{y}}{y}\Big)\Big]
-\frac{2 \alpha x}{(1+\xi x)^{2}}+\frac{\xi^2 x}{(1+\xi x)^2},
\end{eqnarray}
we have the following principal equation which helps to generate various charged models of anisotropic compact relativistic object. It is crucial to integrate Eq.(\ref{28}), because of its non-linearity that could be reformulated for different values of polytropic index $\eta$.
\begin{eqnarray}\label{28}
\frac{\dot{y}}{y}&=&\frac{\mathcal{A}}{4}\left[\frac{3\xi+\xi^{2} x-\alpha x}{1+\xi x}\right]+\frac{\mathcal{B}(1+\xi x)}{4 (8\pi)^{\frac{1}{\eta}}}\left[\frac{3\xi+\xi^2 x-\alpha x}{(1+\xi x)^{2}}\right]^{1+\frac{1}{\eta}}
-\frac{ \alpha x}{4(1+\xi x)}+\frac{\xi}{4},
\end{eqnarray}
and
\begin{eqnarray}\nonumber
\frac{d}{dx}\left(\frac{\dot{y}}{y}\right)&=&-\frac{\mathcal{A}(2\xi^{2}+\alpha)}{4(1+\xi x)^{2}}-
\left[\frac{3\xi+\xi^{2}x-\alpha x}{(1+\xi x)^2}\right]^{\frac{1}{\eta}} \left[\frac{\xi^{3}x-\alpha\xi x+\alpha(1+ \eta)+\xi^{2}(5+2\eta)}{\eta(1+\xi x)^{2}}\right]\label{29}\frac{\mathcal{B}}{4(8\pi)^{\frac{1}{\eta}}}\\
&-&\frac{\alpha}{4(1+\xi x)^{2}}.
\end{eqnarray}
For our compact sphere model, the mass function is given by
\begin{eqnarray}\label{30}
M&=&\frac{1}{4}\Big[\frac{\sqrt{x}\ (2\xi^{3}x-3\alpha-2\xi x\alpha)}{\xi^{2}(1+\xi x)}+\frac{3\ \alpha\  arctan\sqrt{\xi x}}{ \xi^{\frac{5}{2}}}\Big].~~~~~~~~~~~~~~~~~~~~~~~~~~~~~~~~~~~
\end{eqnarray}
The compactness of the star can be expressed by
\begin{eqnarray}\label{30a}
u=\frac{M}{\sqrt{x}}&=&\frac{1}{4}\Big[\frac{ (2\xi^{3}x-3\alpha-2\xi x\alpha)}{\xi^{2}(1+\xi x)}+\frac{3\ \alpha\  arctan\sqrt{\xi x}}{ \xi^{\frac{5}{2}}\sqrt{x}}\Big].~~~~~~~~~~~~~~~~~~~~~~~~~~~~~~~~~~~
\end{eqnarray}
Here, we compare the novelty of our class of exact analytical solutions of the Finch-Skea model with previous variety of works correspond to the four different polytropic index, i.e., $\eta= 1, 2, \frac{2}{3}, \frac{1}{2}$. In the literature, there have been numerous fruitful investigations available with Finch-Skea geometries in the context of $GR$. Very recently, Malaver and Iyer \cite{52v1} have proposed new charged Finch-Skea model for configuration of astrophysical $COs$ by admitting the linear dark energy $EoS$. Dey and Paul \cite{52v2} have studied various notable features of charged anisotropic relativistic solutions for higher dimensional Finch-Skea spacetime by considering the different viable compact star candidates. The impacts of electromagnetic field on the spherically symmetric Finch-Skea star model in the background of two specific choices of polytropic index $\eta=1$ and $2$ have determined by Ratanpal \cite{52v3}. The consequences of static and spherically symmetric anisotropic charged Bardeen spheres have reviewed by indulging the Finch-Skea ansatz obeying the Karmarkar condition \cite{52v4}. In some recent insights \cite{48c,52v5}-\cite{52v8}, several class of exact solutions of anisotropic uncharged Finch-Skea models have been put forward into the account of different mathematical approaches by proposing the observational data of various well-known $COs$. In comparison to these concrete evidences, our results are quite novel and specifically more generic in the background of generalized polytropic $EoS$ rather than the previous finding \cite{52v3}.
{\section{Polytropic models}
The polytropic $EoS$ has an adequate implications in the prediction of astrophysical relativistic compact stars. For instance, Azam and Mardan \cite{52a} and Mardan and Azam \cite{52b} have studied the charged anisotropic polytropes within the framework of generalized polytropic $EoS$. Kippenhahn et al. \cite{52c} have investigated the polytropes with the infinite radius for $\eta=5$. The different properties of uncharged anisotropic stellar relativistic $COs$ with polytropic $EoS$ and $MIT$ $Bag$ model have been proposed in \cite{37}-\cite{38}. Here, we can generate new exact solutions for charged anisotropic compact star, which are physically admissible corresponding to the different values of polytropic index $\eta=\frac{1}{2}, \frac{2}{3}, 1, 2$.
\subsection{Model-1: $\eta=1$}
For $\eta=1$, the generalized polytropic $EoS$ converted into quadratic $EoS$, i.e.,
\begin{equation}\label{31}
P_{r}=\mathcal{A}\rho+\mathcal{B}\rho^{2}.
\end{equation}
Integrating Eq.(\ref{28}) and using above expression, we have
\begin{eqnarray}\nonumber
\ln y&=&\left[\frac{\mathcal{A}(2\xi^{2}+\alpha)+\alpha}{4\xi^{2}}+\frac{\mathcal{B} (\xi^{2}-\alpha)^{2}}{32 \pi \xi^{3}}\right]\ln(1+\xi x)
-\frac{\mathcal{B} (2\xi^{2}+\alpha)}{64 \pi \xi^{3}}
\left[\frac{6\xi^{2}+4\xi^{3}x-3\alpha-4\xi x\alpha}{(1+\xi x)^{2}}\right]\\\label{32}&+&\frac{(\mathcal{A}+1)(\xi^{2}-\alpha)x}{4 \xi}+\mathcal{C},
\end{eqnarray}
where $\mathcal{C}$ is integration constant. In more simplest form this solution can be written as
\begin{equation}\label{33}
y=\mathcal{C}\ (1+\xi x)^{i}\ \exp[L(x)].
\end{equation}
The form of function $L(x)$ and constant $i$ is given as
\begin{eqnarray}\nonumber
L(x)&=&\frac{(\mathcal{A}+1)(\xi^{2}-\alpha)x}{4 \xi}-\frac{\mathcal{B} (2\xi^{2}+\alpha)}{64 \pi \xi^{3}}
\Big[\frac{6\xi^{2}+4\xi^{3}x-3\alpha-4\xi x\alpha}{(1+\xi x)^{2}}\Big],\\
i&=&\frac{\mathcal{A}(2\xi^{2}+\alpha)+\alpha}{4\xi^{2}}+\frac{\mathcal{B} (\xi^{2}-\alpha)^{2}}{32 \pi \xi^{3}}\label{34}.
\end{eqnarray}
By assuming $\mathcal{D}$ as a square of integration constant, the line-element gets the form
\begin{eqnarray}\label{35}
ds^{2}&=&-\mathcal{D}\ (1+\xi r^{2})^{2i}\ \exp[2\ L(r^{2})]\ dt^2+(1+\xi r^{2})\ dr^{2}
+r^{2}\ d\Omega^2.
\end{eqnarray}
Further, for $\alpha=0$, we have the following new uncharged anisotropic compact sphere model
\begin{eqnarray}\nonumber
ds^{2}&=&-\mathcal{D}\ (1+\xi r^{2})^{\mathcal{A} +\frac{\mathcal{B} \xi}{16\pi}}\
\exp\left[\frac{\xi r^{2}}{2}(\mathcal{A}+1)-\frac{\mathcal{B} \xi}{16\pi}\left(\frac{6+4\xi
r^{2}}{1+\xi r^{2}}\right)\right]dt^{2}
+(1+\xi r^{2})\ dr^{2}\\ \label{36}&+&r^2d\Omega^2.
\end{eqnarray}
For $\eta=1$, the matter variable $P_{r}$ and $P_{t}$ as well as pressure anisotropy $\Delta$ are obtained by using the value of $y$ in Eqs.(\ref{11})-(\ref{14}), the new values of matter variables are given in Appendix Eqs.(\ref{55})-(\ref{57}), respectively.
\subsection{Model-2:  $\eta=2$}
For $\eta=2$, the generalized polytropic $EoS$ is written as
\begin{equation}\label{37}
P_{r}=\mathcal{A}\rho+\mathcal{B}\rho^{\frac{3}{2}}.
\end{equation}
From the solution of Eq.(\ref{28}) along with the above choice, we have
\begin{eqnarray}\nonumber
\ln y&=&\left[\frac{\mathcal{A}(2\xi^{2}+\alpha)+\alpha}{4\xi^{2}}\right]\ln(1+\xi x)
+\frac{(\mathcal{A}+1)(\xi^{2}-\alpha)x}{4\xi}\\ \nonumber
&+&\frac{3\mathcal{B}(\xi^{2}-\alpha)\sqrt{2\xi^{2}+\alpha}}{16\sqrt{2\pi}a^{\frac{5}{2}}}
\ln\left[\frac{\sqrt{2\xi^{2}+\alpha}-\sqrt{\xi}\sqrt{3\xi+(\xi^{2}-\alpha)x}}
{\sqrt{2\xi^{2}+\alpha}+\sqrt{\xi}\sqrt{3\xi+(\xi^{2}-\alpha)x}}\right]\\\label{38}&-&\frac{\mathcal{B}}{8 \sqrt{2 \pi}}\left[\frac{\sqrt{3\xi+(\xi^{2}-\alpha) x}\ [2\xi(\alpha-\xi^{2})x+3\alpha]}{\xi^{2}(1+\xi x)}\right]+\mathcal{C},
\end{eqnarray}
where $\mathcal{C}$ indicates an integration constant. Using some algebraic technique, we notice that
\begin{eqnarray}\label{39}
y&=&\mathcal{C}\ (1+\xi x)^{j}\left[\frac{\sqrt{2\xi^{2}+\alpha}-\sqrt{\xi}\sqrt{3\xi+(\xi^{2}-\alpha)x}}
{\sqrt{2\xi^{2}+\alpha}+\sqrt{\xi}\sqrt{3\xi+(\xi^{2}-\beta)x}}\right]^{k}\exp[\mathcal{M}(x)].
\end{eqnarray}
The expression of function $\mathcal{M}(x)$ and constants $j$, $k$ are represented as
\begin{eqnarray}\nonumber
\mathcal{M}(x)&=&\frac{(\mathcal{A}+1)(\xi^{2}-\alpha)x}{4\xi}
-\frac{\mathcal{B}}{8 \sqrt{2 \pi}}\left[\frac{\sqrt{3\xi+(\xi^{2}-\alpha) x}\ [2\xi(\alpha-\xi^{2})x+3\alpha]}{\xi^{2}(1+\xi x)}\right],\\ \nonumber
j&=&\frac{\mathcal{A}(2\xi^{2}+\alpha)+\alpha}{4\xi^{2}}\label{40}, \\
k&=&\frac{3\mathcal{B}(\xi^{2}-\alpha)\sqrt{(2\xi^{2}+\alpha)}}{16\sqrt{2\pi}\xi^{\frac{5}{2}}}.
\end{eqnarray}
By taking $\mathcal{C}^2=\mathcal{D}$, the line-element becomes
\begin{eqnarray}
ds^{2}&=&-\mathcal{D}\ (1+\xi r^{2})^{2j}\times \left[\frac{\sqrt{2\xi^{2}+\alpha}-\sqrt{\xi}\sqrt{3\xi+(\xi^{2}-\alpha)r^{2}}}
{\sqrt{2\xi^{2}+\alpha}+\sqrt{a}\sqrt{3\xi+(\xi^{2}-\alpha)r^{2}}}\right]^{2k }\nonumber \\
&\times& \exp[2\mathcal{M}(r^{2})]\ dt^2+(1+\xi r^{2})\ dr^{2}+r^{2}d\Omega^2\label{41}.
\end{eqnarray}
Here, we can obtain new uncharged anisotropic compact star model for the choice of $\eta=2$
\begin{eqnarray}\nonumber
ds^{2}&=&-\mathcal{D}\ (1+\xi r^{2})^{\mathcal{A}}\left[\frac{\sqrt{2}-\sqrt{3+\xi r^{2}}}{\sqrt{2}+\sqrt{3+\xi r^{2}}
}\right]^{\frac{3\mathcal{B}\sqrt{\xi}}{8\sqrt{\pi}}}\exp\left[\frac{\xi r^{2}}{2}(\mathcal{A}+1)\right.\\ \label{42}
&&\left.+\frac{\mathcal{B} \xi^{\frac{3}{2}}\sqrt{2}\ r^{2}}{4\sqrt{\pi}}\ \left(\frac{\sqrt{3+\xi r^{2}}}{1+\xi^{2}}
\right)\right]\ dt^2+(1+\xi r^{2})\ dr^{2}+r^{2}d\Omega^2.
\end{eqnarray}
For $\eta=2$, the radial and tangential pressures as well as anisotropic factor $\Delta$ can be determined by inserting the outcome of $y$ into Eqs.
(\ref{11})-(\ref{14}), which have distinctively given in Appendix Eqs.(\ref{4c})-(\ref{6c}), respectively.
\subsection{Model-3: for $\eta=2/3$}
For the value of $\eta=2/3$, the generalized polytropic $EoS$ becomes
\begin{equation}\label{43}
P_{r}=\mathcal{A}\rho+\mathcal{B}\rho^{\frac{5}{2}}.
\end{equation}
The solution of Eq.(\ref{28}) for the above consideration takes the form
\begin{eqnarray}\nonumber
\ln y&=&\left[\frac{\mathcal{A}(2\xi^{2}+\alpha)+\alpha}{4\xi^{2}}\right]\ln(1+\xi x)
+\frac{5\ \mathcal{B}\ (\xi^{2}-\alpha)^{3}}{1024\ \sqrt{2}\ \pi^{\frac{3}{2}}\ a^{\frac{7}{2}}\ \sqrt{2\xi^{2}+\alpha}}\\ \nonumber
&\times& \ln\left[\frac{\sqrt{2\xi^{2}+\alpha}-\sqrt{\xi}\sqrt{3\xi+(\xi^{2}-\alpha)x}}
{\sqrt{2\xi^{2}+\alpha}+\sqrt{\xi}\sqrt{3\xi+(\xi^{2}-\alpha)x}}\right]-\frac{\mathcal{B}\ \sqrt{3\xi+(\xi^{2}-\alpha) x}}{1536\ \sqrt{2}\ \pi^{\frac{3}{2}}\ \xi^{3}}\\ \nonumber
&\times&\left[\frac{33(\xi^{2}-\alpha)^{2}}{(1+\xi x)}+\frac{26(\xi^{2}-\alpha)(2\xi^{2}+\alpha)}{(1+\xi x)^{2}}
+\frac{8(2\xi^{2}+\alpha)^{2}}{(1+\xi x)^{3}}\right]\\ \label{44}
&+&\frac{(\mathcal{A}+1)(\xi^{2}-\alpha)x}{4\xi}+\mathcal{C}.
\end{eqnarray}
Here, $\mathcal{C}$ is the constant of integration. The Eq.(\ref{44}) can be rewritten as
\begin{equation}\label{45}
y=\mathcal{D}\ (1+\xi x)^{l}\left[\frac{\sqrt{2\xi^{2}+\alpha}-\sqrt{\xi}\sqrt{3\xi+(\xi^{2}-\alpha)x}}
{\sqrt{2\xi^{2}+\alpha}+\sqrt{\xi}\sqrt{3\xi+(\xi^{2}-\alpha)x}}\right]^{m}\ \exp[N(x)],
\end{equation}
where the function $N(x)$ and constants $l$ and $m$ are
\begin{eqnarray}\nonumber
N(x)&=&\frac{(\mathcal{A}+1)(\xi^{2}-\alpha)x}{4\xi}-\frac{\mathcal{B}\ \sqrt{3\xi+(\xi^{2}-\alpha) x}}{1536\ \sqrt{2}\ \pi^{\frac{3}{2}}\ \xi^{3}}\\ \nonumber
&\times&\Big[\frac{33(\xi^{2}-\alpha)^{2}}{(1+\xi x)}+\frac{26(\xi^{2}-\alpha)(2\xi^{2}+\alpha)}{(1+\xi x)^{2}}
+\frac{8(2\xi^{2}+\alpha)^{2}}{(1+\xi x)^{3}}\Big],\\ \nonumber
l&=&\frac{\mathcal{A}(2\xi^{2}+\alpha)+\alpha}{4\xi^{2}}, \\ \label{46}
m&=&\frac{5\ \mathcal{B}\ (\xi^{2}-\alpha)^{3}}{1024\ \sqrt{2}\ \pi^{\frac{3}{2}}\ a^{\frac{7}{2}}\ \sqrt{2\xi^{2}+\alpha}}.
\end{eqnarray}
Assuming $\mathcal{C}^2=\mathcal{D}$, we can obtain the following line-element
\begin{eqnarray}\nonumber
ds^{2}&=&-\mathcal{D}\ (1+\xi r^{2})^{2l}\times \left[\frac{\sqrt{2\xi^{2}+\alpha}-\sqrt{a}\sqrt{3\xi+(\xi^{2}-\alpha)r^{2}}}
{\sqrt{2\xi^{2}+\alpha}+\sqrt{\xi}\sqrt{3\xi+(\xi^{2}-\alpha)r^{2}}}\right]^{2t} \exp[2N(r^{2})]\ dt^2\\ \label{47}
&&+(1+\xi r^{2})\ dr^{2}+r^{2}\ d\Omega^2.
\end{eqnarray}
For $\eta=\frac{2}{3}$, the uncharged polytropic model is given by the following line-element
\begin{eqnarray}\nonumber
ds^{2}&=&-\mathcal{D}\ (1+\xi r^{2})^{\mathcal{A}}\times \left[\frac{\sqrt{2}-\sqrt{3+\xi r^{2}}}{\sqrt{2}+\sqrt{3+\xi r^{2}}
}\right]^{\frac{5\ \mathcal{B}\ \xi^{\frac{3}{2}}}{2^{11}\ \pi^{\frac{3}{2}}}}\exp\left[\frac{\xi r^{2}}{2}(\mathcal{A}+1)\right.
\left.+\frac{\mathcal{B} \xi^{\frac{3}{2}}\sqrt{(3+\xi r^{2})}}{48(8\pi)^{\frac{3}{2}}(1+\xi r^{2})^3}\right.\\ \label{48}
&&\left.\times\left[117+118\xi r^{2}+33 \xi^{2} r^{4}\right]\right]\ dt^2 \ (1+\xi r^{2})\ dr^{2}+r^{2}\ (d\theta^{2}+\sin^2\theta\  d\phi^{2}).
\end{eqnarray}
For $\eta=2/3$, the radial and tangential pressures and anisotropic pressure $\Delta$ can be found by putting the value of $y$ into Eqs.
(\ref{11})-(\ref{14}), which are given in Appendix Eqs.(\ref{7c})-(\ref{9c}), respectively.
\subsection{Model 4: for $\eta=1/2 $}
For $\eta=\frac{1}{2}$, the generalized polytropic $EoS$ can be obtained as
\begin{equation}\label{49}
P_{r}=\mathcal{A}\rho+\mathcal{B}\rho^{3}.
\end{equation}
We integrate Eq.(\ref{28}) corresponding the value $\eta=\frac{1}{2}$, and get
\begin{eqnarray}\nonumber
\ln y&=&\left[\frac{\mathcal{A}(2\xi^{2}+\alpha)+\alpha}{4\xi^{2}}\right]\ln(1+\xi x)-\frac{\mathcal{B} \xi^{2}}{256\pi^{2}}[2\xi(\alpha-\xi^{2})x+\alpha-4\xi^{2}]\\ \nonumber
&\times&\left[\frac{2\xi^{2}(\alpha-\xi^{2})^{2}x^{2}
+2\xi(\alpha-4\xi^{2})(\alpha-\xi^{2})x\alpha^{2}-2\xi^{2}\alpha+10\xi^{4}}{4\xi^{4} (1+\xi x)^4}\right]\\ \label{50}
&+&\frac{(\mathcal{A}+1)(\xi^{2}-\alpha)x}{4 \xi}+\mathcal{C},
\end{eqnarray}
where $\mathcal{C}$ is the constant of integration. We can rewrite the above form in the following manner
\begin{equation}\label{51}
y=\mathcal{C}\ (1+\xi x)^{n}\ \exp[O(x)].
\end{equation}
The formulation of variable $O(x)$ and constant $n$ is described as
\begin{eqnarray}\nonumber
O(x)&=&\frac{(\mathcal{A}+1)(\xi^{2}-\alpha)x}{4 \xi}-\frac{\mathcal{B} \xi^{2}}{256\pi^{2}}[2\xi(\alpha-\xi^{2})x+\alpha-4\xi^{2}]\\ \nonumber
&\times&\left[\frac{2\xi^{2}(\alpha-\xi^{2})^{2}x^{2}
+2\xi(\alpha-4\xi^{2})(\alpha-\xi^{2})x\alpha^{2}-2\xi^{2}\alpha+10\xi^{4}}{4\xi^{4} (1+\xi x)^4}\right]+\mathcal{C},\\ \label{52}
n&=&\frac{\mathcal{A}(2\xi^{2}+\alpha)+\alpha}{4\xi^{2}}.
\end{eqnarray}
We assume $\mathcal{C}^{2}=\mathcal{D}$, then metric has the following form
\begin{eqnarray}\label{53}
ds^{2}&=&-\mathcal{D}\ (1+\xi r^{2})^{2n}\exp[2\ O(r^{2})]\ dt^2+(1+\xi r^{2})\ dr^{2}+r^{2}\ d\Omega^2.
\end{eqnarray}
Further, we can get the following line-element for an uncharged polytropic model by considering $\eta=\frac{1}{2}$,
\begin{eqnarray}\nonumber
ds^{2}&=&-\mathcal{D}\ (1+\xi r^{2})^{\mathcal{A}}\ \exp\left[\frac{\xi r^{2}}{2}(\mathcal{A}+1)+\frac{\mathcal{B} \xi^2}{128 \pi^{2}}\right.\\ \label{54}
&&\left.\times\left(\frac{\xi^{3} r^{6}+6 \xi^2 r^{4}+13\xi r^{2}+10}{(1+\xi r^{2})^4}\right)\right]\ dt^2+(1+\xi r^{2})\ dr^{2}+\ r^{2}\ d\Omega^2.
\end{eqnarray}
For $\eta=1/2$, the radial and tangential pressure as well as anisotropic factor $\Delta$ can be explored by putting the result of $y$ into Eqs. (\ref{11})-(\ref{14}), which are given in Appendix Eqs.(\ref{10c})-(\ref{12c}), respectively. More specifically, the results of charged anisotropic compact star models inspired by a generalized polytropic $EoS$ can be retrieved into linear, quadratic and polytropic $EoS$. When we set $\mathcal{B}=0$, $\eta=1$ and $\mathcal{A}=0$ the generalized polytropic $EoS$ recovered as linear, quadratic and polytropic $EoS$, respectively. In the gravitational and astrophysical perspectives, these $EoSs$ have observed variety of implications; for instance, the linear $EoS$ has described the dust fluid ($\mathcal{A}=0$), radiation matter ($\mathcal{A}=\frac{1}{3}$) \cite{43}, and it has elaborated the quark made star configuration \cite{30,31}. The quadratic $EoS$ has proposed the various stellar astrophysical compact models like charged relativistic strange star and quark star \cite{33,34}. Moreover, the polytropic $EoS$ has inspected the numerous relativistic astrophysical $COs$ like white dwarfs, brown dwarfs, neutron stars, whereas it has explained the early and late time Universe with positive and negative values of the polytropic index \cite{41,52c0}.
{\section{Physical attributes of polytropic models}
We have observed the physical stability constraints for an under developed polytropes models via generalized polytropic $EoS$ in the spacetime geometry of Finch and Skea. It is predicted that the proposed ansatz is physically reliable as it is continuous everywhere and potentially stable inside the sphere. To choice the gravitational potential as seed ansatz is ever being a challenged among the theoretical physicists and cosmologists for the validation of maximum essential physical requirements of the system. In favor of the purposed ansatz, there have been made wide range of substantial explorations to check the sustainability and regularity of the cosmic objects. It has admitted a key role in number of aspects by explaining the stellar relativistic astrophysical as well as cosmological consequences. For instance, this form has contemplated as seed ansatz to intimate that Universe is in a phase of accelerated expansion by imposing the dark energy $EoS$ \cite{52v1}. In the usual four and higher dimensions the solution of Finch-Skea has disclosed to understand some physical implications at extreme conditions of charged compact stellar models \cite{52v2}. This geometry has validated various stability tests with physical variables of the astrophysical $COs$ under by the influence of polytropic $EoS$ \cite{52v3, 52v8}. In particular, the Finch-Skea model as an interior spacetime has manifested several viable features with an exterior solution of Bardeen black hole which corresponds to the magnetic monopole of gravitationally collapsing remnants \cite{52v4,52c1,52c2}. Moreover, it has successfully employed to endorse the observational constraints of various type of astrophysical compact stars under by the domination of various physical conditions \cite{48c,52v5}-\cite{52v6}. Apart of these inspirational outcomes, in most of the recent works the finch and skea geometry has positively revealed the stellar relativistic astrophysical implications with contribution of the Adler and buchdahl spacetimes \cite{27a,50p,52v4,52c1}-\cite{52c3}. These spacetimes have procured the solutions as like to Finch and skea model by obeying the various astrophysical bounds of the compact stars. Therefore, it fulfills the following conditions:
\begin{eqnarray}\nonumber
e^{2\lambda(0)}&=&\rm{constant}, \\ \nonumber
e^{2\psi(0)}&=&1,
\end{eqnarray}
\begin{figure}
\begin{center}
\includegraphics[width=.47\linewidth, height=2.4in]{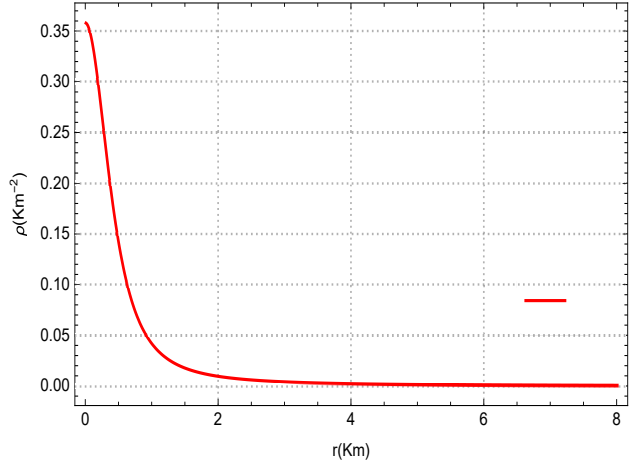}
\caption{Evolutionary behavior of matter density as a function of radial coordinate $r$; for $\xi=3$ and $\alpha=1.15$.}
\end{center}
\end{figure}

\begin{figure}
\begin{center}
\includegraphics[width=.46\linewidth, height=2.4in]{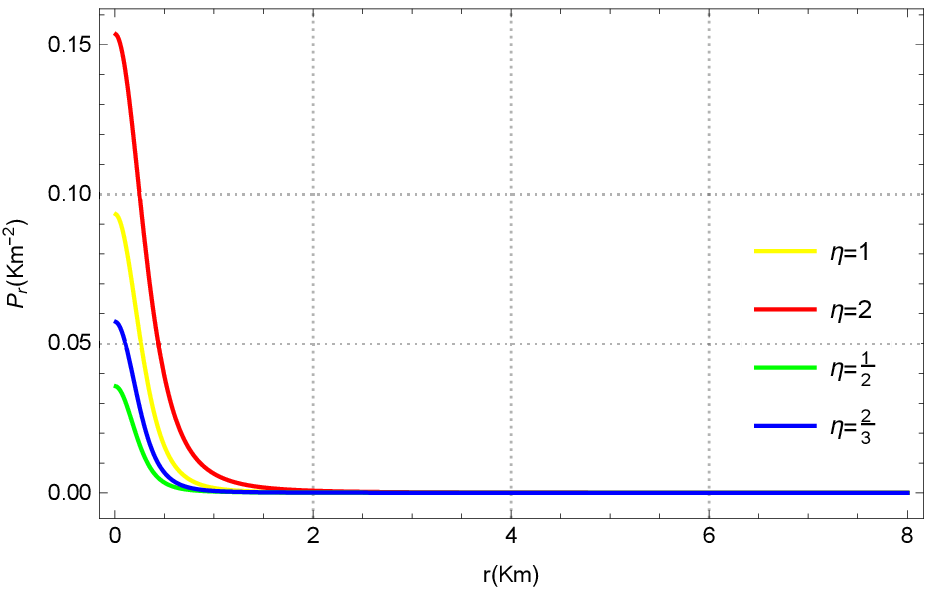}
\includegraphics[width=.46\linewidth, height=2.4in]{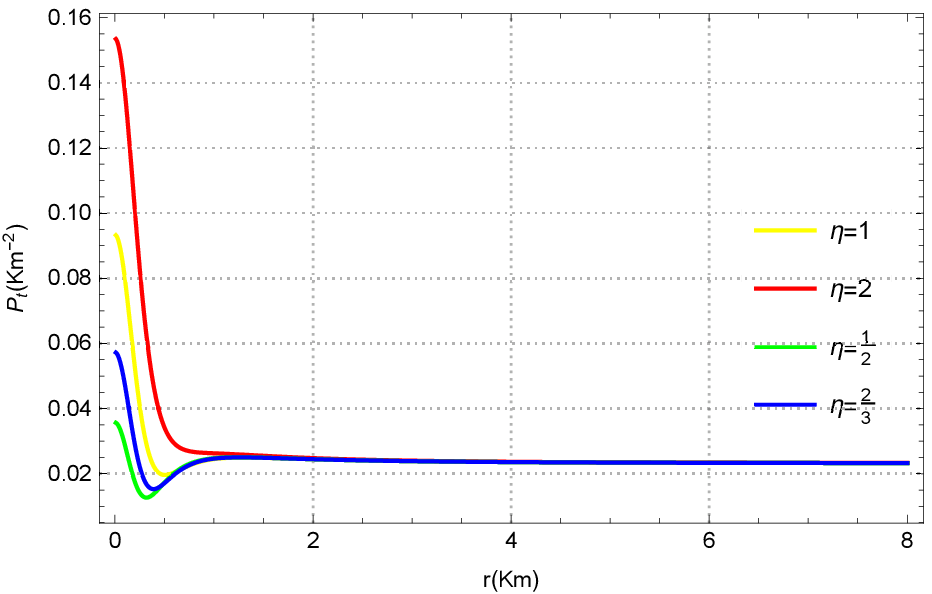}
\caption{Physical evolution of radial and tangential pressures vs $r$; for different choices of $\eta=1, 2, \frac{1}{2}, \frac{2}{3}$, $\xi=3, \mathcal{A}=0.01, \mathcal{B}=0.7$ and $\alpha=1.15$.}
\end{center}
\end{figure}
and $(e^{2\psi(r)})'=(e^{2\lambda(r)})'=0 $ when $r=0$. In this analysis, we study the graphical impact of various polytropic models under numerous physical properties like, matter density, pressure ($P_{r}$ and $P_{t}$), pressure anisotropy, mass function, electric field, speed of sound, compactness, electric charge and adiabatic index. We observe the physical viability of matter density for the evolution of compact sphere model in Fig. \textbf{1}. It can be seen from Fig. \textbf{1}, that the behavior of matter density is non-negative as well as non-singular throughout the whole interior region of the star. It is also noticed here that the matter density is maximum near the center as compared to the surface of the star. This characteristic suggests that the presented model predict an ultra compact relativistic object. In Fig. \textbf{2}, we depict the evolutionary nature of radial and tangential pressure for the modeling of stellar interior compact stars. One can transparently see from Fig. \textbf{2}, that the radial and tangential pressures possess positive evolution as they remain regular (finite) and non-singular at each point inside the compact stellar body. It is also evident here that both material quantities are increasing from center and are decreasing gently towards the boundary surface.

In formation of static and spherically symmetric anisotropic $COs$, we described here some viable consequences of Einstein equations of motion with their standard physical features. These consequences can be obtained in absence of pressure, isotropic source and anisotropic fluid. Nevertheless, there are many vigorous probes that steep excessive compact relativistic stars are not formed of isotropic sources. In certain bounds, the structures with unpredictable physical implications are formed, for the illustration of pressure anisotropy. In Particular, we briefly review the anisotropic matter field via energy-momentum tensor that characterizing the physical composition of stellar interior $CO$. Indeed, anisotropic matter field is an exceptional exotic choice for modeling of stellar $COs$ like, white dwarfs, quark made stars, neutron stars, pulsars, etc. Keeping in view, Jeans \cite{52d} was the first theorist who reviewed the effects of anisotropic pressure on the self-gravitating relativistic dense star model. Moreover, it has been revealed that the pressure anisotropy may appear within an interior configuration of the compact stars due to various properties like, superluminal fluid \cite{20,21,52c}, phase periods \cite{52d} and electromagnetic field \cite{52e}. Based on these vital facts, we examine the profile of pressure anisotropy for our proposed polytropic models in Fig. \textbf{3}. It can be easily seen from Fig. \textbf{3}, that the evolution of pressure anisotropy is positive as well as regular (finite) within the distribution of the star but at some limiting points it is slightly deviated from the standard position because it might be due to exotic nature of the fluid. The pressure anisotropy sharply proceeds from center towards the boundary surface of the sphere where $r=1$. In addition, this trend indicates that the pressure anisotropy have repulsive nature, i.e., $P_{r}<P_{t}$. Actually, this characteristic identifies that an interior core of the star is made up with high density, which is due to increase of pressure anisotropy. As a result, we can reflect that our proposed models of spherically dense compact stars are physically admissible as well as potentially stable.
\begin{figure}
\begin{center}
\includegraphics[width=.46\linewidth, height=2.4in]{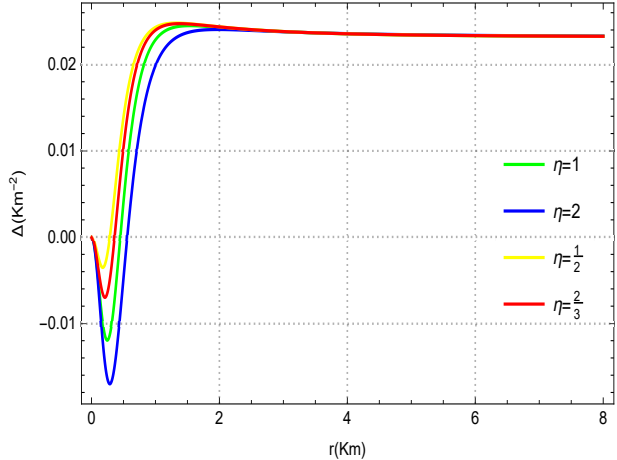}
\caption{Variational nature of pressure anisotropy versus $r$ for various values of $\eta=1, 2, \frac{1}{2}, \frac{2}{3}$, $\xi=3, \mathcal{A}=0.01, \mathcal{B}=0.7$ and $\alpha=1.15$.}
\end{center}
\end{figure}
\begin{figure}
\begin{center}
\includegraphics[width=.46\linewidth, height=2.4in]{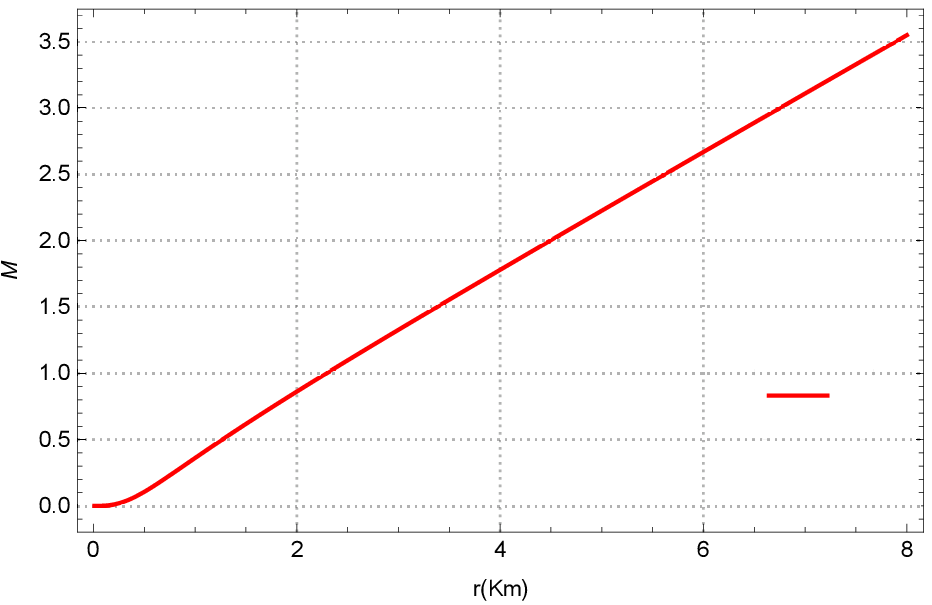}
\includegraphics[width=.46\linewidth, height=2.4in]{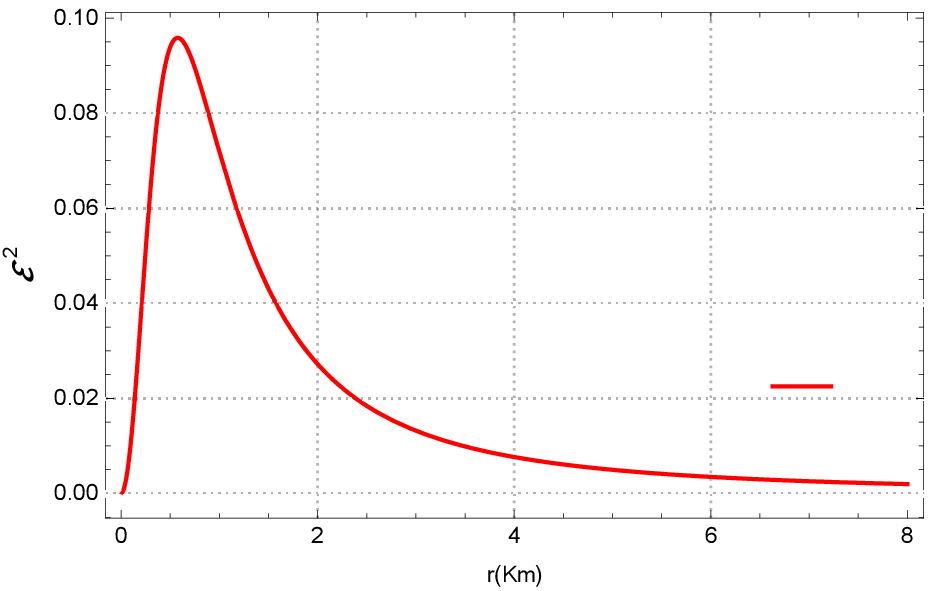}
\caption{Physical variation of effective mass function (left plot) and electromagnetic field (right plot) w.r.t. radial distance $r$.}
\end{center}
\end{figure}
In Fig. \textbf{4}, we analyze the physical evolution of effective mass function as well as electromagnetic field for our compact star model. The behavior of effective mass within an entire region of the star is non-negative as it is continuous (regular) at every point. Also, it is rapidly progressive from center towards the boundary of the sphere and attains maximum value at $r\sim1.5$. From these features, we can predict that our proposed star model is able to develop an ultra dense relativistic star as predicted in Ref. \cite{50d}. Moreover, one can notice from right plot (Fig. \textbf{4}), that the electromagnetic field is positive at every point inside the sphere as it is continuous (regular) in everywhere. Also, it possess maximum evolutionary nature near the core of the star and steadily decreases towards the boundary surface with the increase of radial coordinate $r$.
\begin{figure}
\begin{center}
\includegraphics[width=.44\linewidth, height=2.1in]{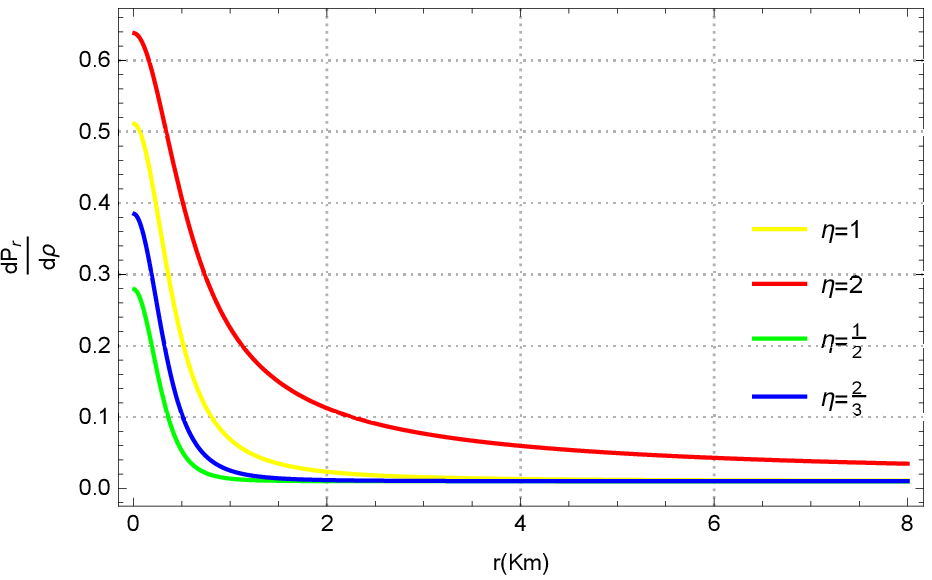}
\includegraphics[width=.44\linewidth, height=2.1in]{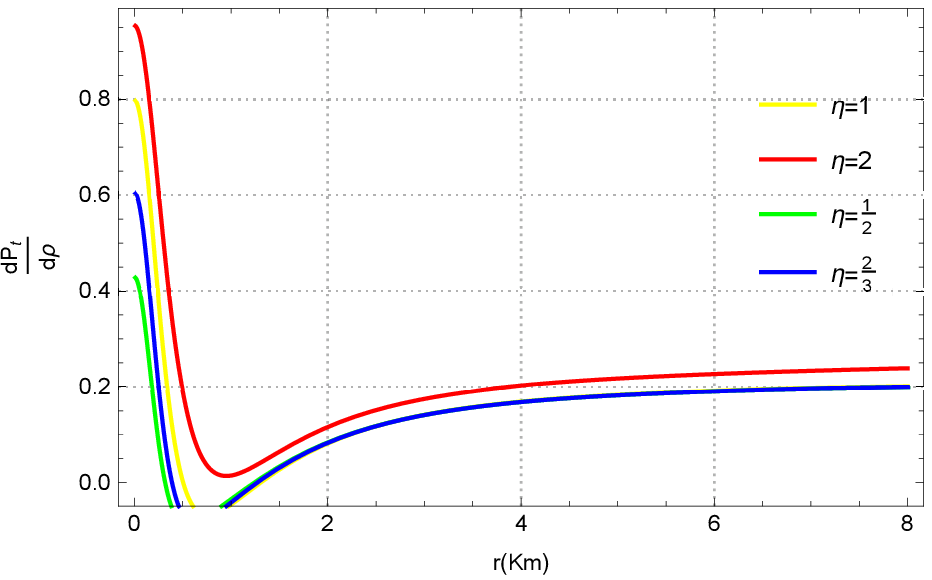}
\caption{Physical variation of sound speed $\frac{dP_{r}}{d\rho}$ (left plot) and $\frac{dP_{t}}{d\rho}$ (right plot) versus $r$; for certain values of $\eta=1, 2, \frac{1}{2}, \frac{2}{3}$, $\xi=3, \mathcal{A}=0.01, \mathcal{B}=0.7$ and $\alpha=1.15$.}
\end{center}
\end{figure}

We study the evolutionary nature of sound speed ($\frac{dP_{r}}{d\rho}$ and $\frac{dP_{t}}{d\rho}$) for our polytropic models of compact stars analogous to various choices of $\eta$ in Fig. \textbf{5}. It is evident to observe from Fig. \textbf{5}, that the trend of all curves of sound speed (left plot) with respect to their polytropic models suggest positive nature at any interior point of the star. Moreover, it would be very interesting to mention here that for our proposed polytropic models sound speed (left plot) remains within the certain limit and does not cross the sufficient value, i.e., $0<\frac{d\mathrm{P_{r}}}{d\rho}\leq1$. On the other hand, the behavior of tangential speed of sound remains positive throughout the whole configuration of the star for all polytropic models and it does not cross the sufficient bound, i.e., $0<\frac{d\mathrm{P_{t}}}{d\rho}\leq1$ (see Fig. \textbf{5} (right plot)). Moreover, the nature of tangential sound speed regular (finite) at every interior point and it is gradually decreases from center outward towards the boundary surface of the star. These all features of sound speed once again strongly recommend that our compact star models are stable as well as compatible with the prior result of causality condition depicted in \cite{6}.

\begin{figure}
\begin{center}
\includegraphics[width=.46\linewidth, height=2.4in]{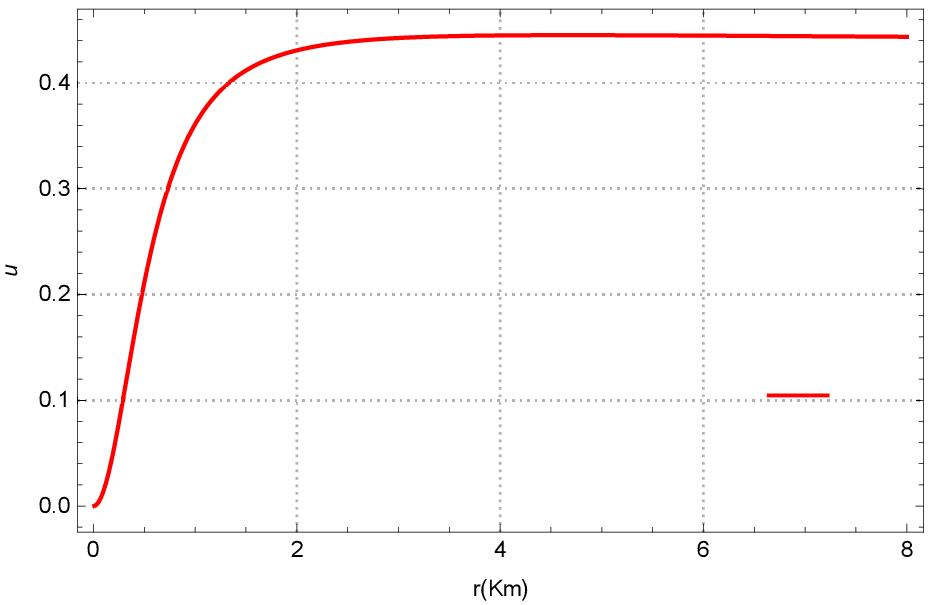}
\includegraphics[width=.46\linewidth, height=2.4in]{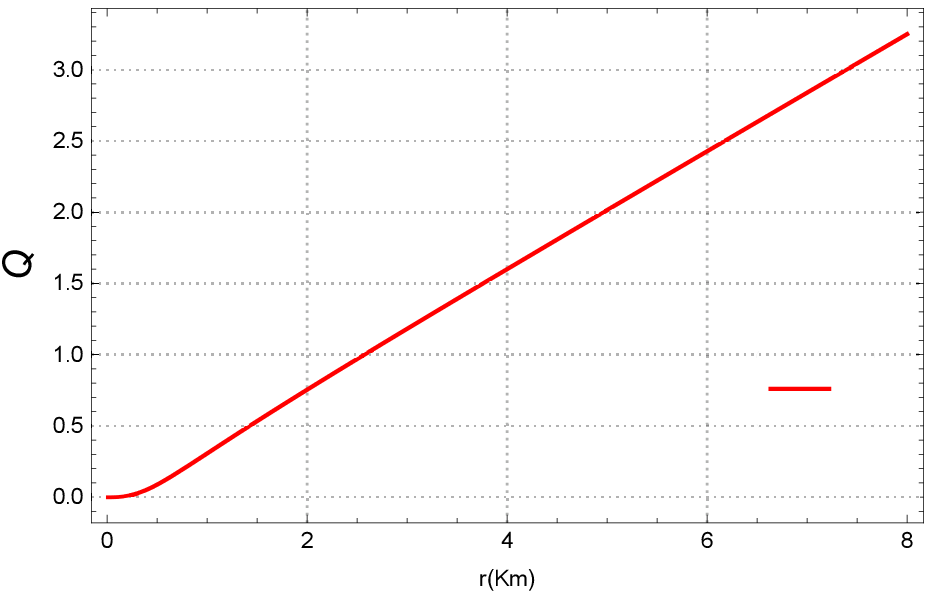}
\caption{Physical variation of compactness (left plot) and total electric charge (right plot) versus radial distance $r$.}
\end{center}
\end{figure}

\begin{table}[ht]
\caption{Numerically estimated value of total electric charge and compactness parameter corresponds to the arbitrary radius and mass of the compact star}
\begin{center}
\begin{tabular}{|c|c|c|c|c|c|c|c|c|}
\hline$r_{b}$&\textbf{$\textit{M}(\textit{M}_\odot)$}&\textbf{$\textit{M}$}&\textbf{$Q$}&\textbf{$\frac{Q}{r_{b}}$}&\textbf{$\frac{Q^{2}}{r^{2}_{b}}$}&\textbf{$\frac{\textit{M}}{r_{b}}$}&\textbf{$\frac{2\textit{M}}{r_{b}}$}\\
     $(Km)$  &                     &$(Km)$&($\textit{Coulomb}$ $\textit{C}$)       &                          &                                  & $<\frac{4}{9}$          & $<\frac{8}{9}$
\\\hline 8 & 2.40661&3.54976            &3.7880$\times 10^{20}$    &0.406133                  &0.164944                          & 0.443720  &0.887441
\\\hline
\end{tabular}
\end{center}
\end{table}
A quite realistic relation between the compactness and an electric charge has ever remained pivotal regarding the non-violation of its stable configuration in the modeling of $COs$. For this assessment, the pioneering study of compactness factor for static and spherically symmetric perfect fluid sphere was proposed by Buchdahl \cite{52e1}. According to Buchdahl, the maximum allowable ratio of mass-radius for a $CO$ can be less than $\frac{4}{9}$ i.e., ($\frac{2\textit{M}}{r_{b}}<\frac{8}{9}$). Mak et al. \cite{52e2} and Andr$\acute{e}$asson \cite{52e3} have generalized such condition for an interior charged compact star models. The term $\frac{\textit{M}}{r_{b}}$ is considered as compactness factor which categorizes the astrophysical $COs$ as normal star; $\frac{\textit{M}}{r_{b}}\sim 10^{-5}$, white dwarf; $\frac{\textit{M}}{r_{b}}\sim 10^{-3}$, neutron star; $0.1<\frac{\textit{M}}{r_{b}}<0.25$, ultra dense $CO$ ; $0.25<\frac{\textit{M}}{r_{b}}<0.50$ and black hole; $\frac{\textit{M}}{r_{b}}=0.50$ \cite{52e4}. It would be very interesting to mention here that the numerically estimated value of an electric charge and compactness for our compact star model is shown in Table \textbf{1}. One can easily inspect from the given Table \textbf{1}, that the numerically estimated ratio of charge to the radius for the given compactness is in allowable limit and does not violate the Buchdahl condition for charged $CO$. Despite of this allowable limit, there is another upper limit that should necessarily be satisfied the maximum allowed compactness and charge-radius ratio, i.e., ($\frac{Q^{2}}{r^{2}_{b}}<\frac{2\textit{M}}{r_{b}}<\frac{8}{9}$). This whole discussion vigorously confirmed that our compact star model fulfilled the necessary Buchdahl condition for charged $CO$ as well as might be able to predict a ultra dense $CO$. There is another interesting comparison presented here with previous momentous works regarding the computed amount of total electric charge in the coulomb unit given in Table \textbf{1}. We observed that the total charge on the surface in coulomb unit is 3.7880$\times 10^{20}$ [C] for the proposed star whose arbitrary mass and radius are ${M}_\odot$=3.54976 [km] and $r_{b}=8$ [km]. As investigation of $f(R,T)$ gravity in the background of two geometrical techniques analyzed by Maurya et al. \cite{50p}, in which they estimated the maximum amount of electric charge as 2.46693$\times 10^{20}$ [C] and 2.58885$\times 10^{20}$ [C] for two different charged compact star models. In the realm of $GR$ the same author and his collaborators \cite{50n,52v} have shown the maximum amount of electric charge in coulomb unit for anisotropic strange star candidates as 4.04378$\times 10^{19}$ [C] whose radius is 8 [km], and 1.638631$\times 10^{20}$ [C] whose radius is 7.1 [km], respectively, as well as they estimated 7.5317$\times 10^{19}$ [C] for PSR J1664-2230 and 9.8951$\times 10^{19}$ [C] for PSR 1937 + 21. From this detail analysis, our computed amount of electric charge on the surface in coulomb unit is more better comparatively to these discussed results.

Moreover, the physical behavior of compactness and an electric charge for our compact star model is shown in Fig. \textbf{6}. It can be clearly observed from Fig. \textbf{6}, that the trend of compactness (left plot) and electric charge (right plot) suggest positive evolution at every point of the stellar interior as they are non-singular within the entire region of the star. The compactness and electric charge become increasing function of radial coordinate $r$ and they attain maximum value at the boundary sphere. Also, it is mentioned here that the Buchdahl limit for charged $CO$ remains in good agreement and does not violate the necessary condition. In spite of these facts, the mass-radius relation yields $0.25<\frac{\textit{M}}{r_{b}}<0.50$, which correspond to stable neutron star and ultra dense compact star.
\begin{table}[ht]
\caption{The numerical value of adiabatic index $\Gamma(0)$ along with critical value $\Gamma$(crit) for the proposed polytropic models are given as}
\begin{center}
\begin{tabular}{|c|c|c|c|c|c|c|c|c|}
\hline$\Gamma(0)$      &$\Gamma(0)$        &$\Gamma(0)$        &$\Gamma(0)$       &$\Gamma$(crit)\\
     (\textbf{Model-1})& (\textbf{Model-2})&(\textbf{Model-3}) &(\textbf{Model-4})&
\\\hline        2.47297 &       2.12667      &     3.07882&2.79126&    1.73479
\\\hline
\end{tabular}
\end{center}
\end{table}
\begin{figure}
\begin{center}
\includegraphics[width=.46\linewidth, height=2.4in]{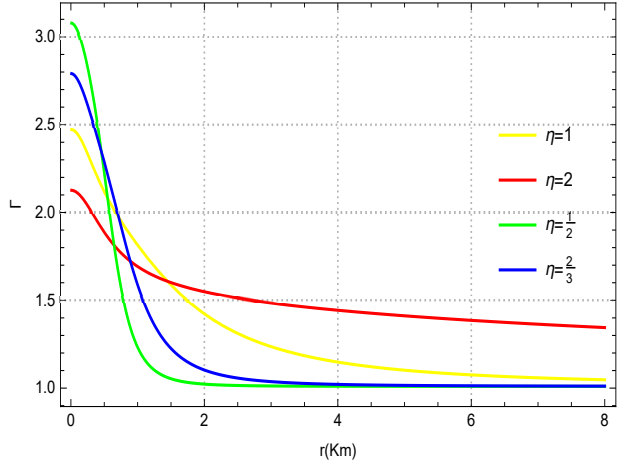}
\caption{Physical behavior of adiabatic index $\Gamma$ vs radial distance $r$.}
\end{center}
\end{figure}

On the contrary of above discussed stability criterion, another most prominent condition for ensuring the stability equilibrium of the model is adiabatic index. The term adiabatic index is referred as the ratio of two specific heats \cite{52e5}. To check the consistency of the model, the relativistic adiabatic index $\Gamma$ in the radial direction plays a prime role. Here, the radial direction is vital, because, under the influence of anisotropy the system is reformed from the spherically symmetric one in the radial direction only for restraining gravitational collapse. In some former papers \cite{52e6}-\cite{52e7}, the detail analysis has been given in different mathematical mechanisms to compute the relativistic adiabatic index along with critical value to ensure the stability of the stellar compact models. The relativistic adiabatic index can be expressed as
\begin{eqnarray}\label{53a}
\Gamma=\frac{\rho+P_{r}}{P_{r}}\frac{dP_{r}}{d\rho}.
\end{eqnarray}
One can affirm that, for stability of the system becomes $\Gamma\geq\Gamma_{crit}$. Whereas, the critical value of adiabatic relativistic index can be evaluated by this expression
\begin{eqnarray}\label{53b}
\Gamma_{crit}=\frac{4}{3}+\frac{19u}{21}.
\end{eqnarray}
In Fig. \textbf{7}, we plot the physical behavior of relativistic adiabatic index $\Gamma$ for different polytropic models. It can be eminently seen from Fig. \textbf{7}, that all curves of the adiabatic relativistic index corresponding to various polytropic models suggest positive evolution within the stellar object as well as gradually decreases from center outward towards the boundary of the sphere. Also, these polytropic models confirm the stability of our stellar $CO$ by satisfying the relativistic condition as $\Gamma>\frac{4}{3}$ (see Fig. \textbf{7}). As a result, we can firmly say that the suggested models are potentially well behaved and being able to model of $CO$. Moreover, we computed the numerical values of $\Gamma$ along with critical value of adiabatic relativistic index shown in Table \textbf{2}. From these computed numerical values, one can understand that the condition (\ref{53b}) is once again ensured the stability equilibrium of our models.
\section{Discussion}
Inspirited from an under evolved polytropic models via generalized polytropic $EoS$ in the Finch and Skea geometry, the formation of stellar relativistic $CO$ might be quite conducive. It has been observed in Ref.\cite{52f} that a non-linear $EoS$ has possessed the features of dark matter and dark energy which is helpful to explain inflationary the phases of our Universe. In this work, we have generated new exact analytical solutions for static and spherically symmetric metric containing the properties of anisotropic matter. The obtained solutions of generalized polytropic $EoS$ are constructed to model charged anisotropic compact stars corresponding to Finch-Skea spacetime. We have examined graphically that models are physically admissible as well as potentially sustainable. It is found that the gravitational potential and matter quantities are non-singular, as these are well-consistent inside the celestial sphere. For $\eta=1$, we can simply concise our solution to the particular case which recommended by Feroze and Siddique \cite{33}. These solutions may be fruitful in the modeling of relativistic $COs$ and consider particular matter distribution due to such kind of generalized polytropic $EoS$. It might be an interesting work to differentiate acquired solutions to manifest astrophysical $COs$ such as $SAX J 1808.4$-$3658$ for charge and uncharge as studied by Mafa Takisa an Maharaj \cite{31} and Dey et al. \cite{53}. It would be very fine tune model to construct the charged relativistic astrophysical compact sphere which is made up of high density. Also, we can revert our results into new uncharged anisotropic solutions by vanishing electric charge. This work reveals that Finch-Skea spacetime is physically suitable for the formation of astronomical $COs$. Form this understanding, we provide the following thorough authentication regarding our proposed solutions via the graphical analysis:
\begin{itemize}
\item Our results of matter variables ($\rho$, $P_{r}$ and $P_{t}$) have confirmed that they are regular and well behaved throughout the whole configuration of stellar object as shown in Figs. \textbf{1} and \textbf{2}. The non-negative and monotonically decreasing nature of such variables can be evidently seen in the graphical plots. Furthermore, we have noticed that near the central core of the star, the nature of these variables is gradually increasing and then sharply vanishing towards the boundary of the star where $r$ is maximum. These implications ultimately suggest that our models is ultra dense relativistic compact stars.

\item From Fig. \textbf{3}, we have clearly observed that the anisotropy of pressure possessed evolutionary nature and continuous at every point inside the stellar body, also it is observed that the anisotropy of pressure have repulsive trend, i.e., $P_{r}<P_{t}$. This behavior indicates that our models are well-consistent and can predict an ultra dense $CO$.

\item The graphical behavior as observed in Fig. \textbf{4}, shows that mass function (left plot) is rapidly increasing from center to the boundary of the star, which further endorse that the present model is viable and potentially stable. We have also studied the physical evolution of electromagnetic field for our compact star model in Fig. \textbf{4} (right plot). The electromagnetic field have non-negative evolution at every point inside the sphere as it is regular everywhere. It is a monotonically decreasing function of radial coordinate $r$ and vanishes at the boundary of the star.
\item Despite of these concrete evidences, we have investigated the physical impact of causality condition ($\frac{d\mathrm{P_{r}}}{d\rho}$ and $\frac{d\mathrm{P_{t}}}{d\rho}$) on our compact star as shown in Fig. \textbf{5}. The evolution of each curve of causality condition ($\frac{d\mathrm{P_{r}}}{d\rho}$) remains in the appropriate range and satisfy $0<\frac{d\mathrm{P_{r}}}{d\rho}\leq1$. Also, the transverse sound speed behaves regular nature at each point inside the star as well as non-negative throughout the entire region. Beside of this characteristics, the all trends of the transverse sound of speed with respect to the various choices of $\eta$ show good agrement and doest not breach the sufficient limit as $0<\frac{d\mathrm{P_{t}}}{d\rho}\leq1$. The causality condition once again strongly agreed that the presented models of compact star are stable.
\item The variation of compactness and electric charge are plotted against $r$ in Fig. \textbf{6}. One can transparently see from Fig. \textbf{6}, that the nature of both plots (left and right) is continuous at all points in the interior of the star. The figure also indicates that the compactness and electric charge are monotonically increasing function of $r$ and they attained maximum amount at $r\sim1.5$. Also, the Buchdahl condition for charged $CO$ remained in an allowable range and satisfy the necessary bound.
\item We have depicted the physical behavior of adiabatic relativistic index $\Gamma$ for the proposed polytropic models in Fig. \textbf{7}. The all curves of $\Gamma$ corresponding to various choices of $\eta$ propose good stable agreement at every point within the compact sphere. It is also confirmed from Table \textbf{2}, the value of adiabatic relativistic index for all models remain greater than critical value of adiabatic index throughout the celestial system and it represents a stable equilibrium position to compose a viable $CO$.
\end{itemize}
Conclusively, the obtained features of the generalized polytropic $EoS$ along with Finch-Skea ansatz are quite conducive and potentially stable in hydrostatic equilibrium for describing the gravitational aspects of the astrophysical $COs$. There have several physical conditions satisfied via different criterions, viz. mass function, electric field, causality bounds, Buchdahl condition with charge to radius ratio and adiabatic index along with critical value. Particulary, our estimated value of compactness nearly close to Buchdahl limit, and it has evidently justified with upper bound of charge-radius ratio for charged $CO$ (see Table \textbf{1}). It is another intriguing aspect to note here that our observed amount of total electric charge at the boundary in the coulomb unit as 3.7880$\times 10^{20}$ [C] correspond to the arbitrary mass ${M}_\odot$=3.54976 [km] and radius $r_{b}=8$ [km], which is more better than in comparison of the previous work proposed in \cite{50n,50p,52v}. We have once again revitalized our results to ensure the stability of our models by inspecting the relativistic adiabatic index along with critical value. Both stability conditions are in good agreement and please the sufficient relativistic physical bounds (see Table \textbf{2}). These attributes have vigorously confirmed the maximum gravitational stability constraints for composing the viable ultra dense compact star model and neutron star. Such consequences in presence of electric charge are exclusively novel and more generic in the framework of generalized polytropic $EoS$ alternately to preceding works \cite{52v3,52v8}, which were deduced only for the limited frameworks.
\section*{Appendix}
The outcomes of pressure ($P_{r}$ and $P_{t}$) and anisotropy of pressure ($\Delta$) analogous to different choices of polytropic index like, $\eta=1, 2, \frac{2}{3}, \frac{1}{2}$ are given below
\begin{itemize}
\item For $\eta=1$
\begin{equation}\label{55}
P_{r} = \frac{\mathcal{A}}{8\pi} \Big(\frac{3\xi+\xi^2 x-\alpha x}{(1+\xi x)^{2}}\Big)+\frac{\xi}{8\pi}\Big(\frac{3\xi+\xi^2 x-\alpha x}{(1+\xi x)^{2}}\Big)^{2},
\end{equation}
\begin{eqnarray}\nonumber
P_{t}&=&\mathcal{A}\ \Big[\frac{3\xi+\xi^{2}x-\alpha x}{8\pi(1+\xi x)^{2}}\Big]+\xi\Big[\frac{3\xi
+\xi^{2}x-\alpha x}{8\pi(1+\xi x)^{2}}\Big]^{2}+\frac{x}{2\pi(1+\xi x)} \Big[\frac{i(i-1)\xi^{2}}{(1+\xi x)^{2}}+\frac{2\ i\ \xi\ \dot{L}(x)}{1+\xi x}\\\label{56}
&+&\dot{L}(x)^{2}+\ddot{L}(x)\Big]-\frac{\xi x}{4\pi(1+\xi x)^{2}}\Big[\frac{i\ \xi}{1+\xi x}+\dot{L}(x)\Big]+ \frac{\xi^{2}x}{8\pi(1+\xi x)^{2}}-\frac{\alpha x}{4\pi(1+\xi x)},
\end{eqnarray}
\begin{eqnarray}\nonumber
\Delta&=&\frac{x}{2\pi(1+\xi x)}\Big[\frac{i(i-1)\xi^{2}}{(1+\xi x)^{2}}+\frac{2\ i\ \xi\ \dot{L}(x)}{1+\xi x}+\dot{L}(x)^{2}+\ddot{L}(x)\Big]
-\frac{\xi x}{4\pi(1+\xi x)^{2}}\Big[\frac{i\ \xi}{1+\xi x}\\ \label{57}&+&\dot{L}(x)\Big]+ \frac{\xi^{2}x}{8\pi(1+\xi x)^{2}}
-\frac{\alpha x}{4\pi(1+\xi x)}.
\end{eqnarray}
Here, the derivative w.r.t variable $x$ is represented with dot sign, whereas the function $\dot{L}(x)$ and $\ddot{L}(x)$ are illustrated as
\begin{eqnarray}\nonumber
\dot{L}(x)&=&\frac{(\xi^{2}-\alpha)}{4a}\ (\mathcal{A}+1)+ \frac{\mathcal{B}\ (2\xi^{2}+
\alpha)}{32\pi\xi^{2}}\left[\frac{4\xi^{2}+2\xi^{3} x-2\xi x\alpha-\alpha}{(1+\xi x)^{3}}\right],\\\nonumber
\ddot{L}(x)&=&-\frac{\mathcal{B}\ (2\xi^{2}+\alpha)}{32 \pi \xi}\left[\frac{10\xi^{2}+4\xi^{3} x-4\xi x\alpha-\alpha}{(1+\xi x)^{4}}\right].
\end{eqnarray}
\item For $\eta=2$
\begin{eqnarray}\label{4c}
P_{r}&=&\mathcal{A} \Big[\frac{3\xi+\xi^2 x-\alpha x}{8\pi(1+\xi x)^{2}}\Big]+\mathcal{B} \Big[\frac{3\xi+\xi^2 x-\alpha x}{8\pi(1+\xi x)^{2}}\Big]^{\frac{3}{2}},\\\nonumber
P_{t}&=&\mathcal{A}\ \Big[\frac{3\xi+\xi^{2}x-\alpha x}{8\pi(1+\xi x)^{2}}\Big]+
\mathcal{B}\Big[\frac{3\xi+\xi^{2}x-\alpha x}{8\pi(1+\xi x)^{2}}\Big]^{\frac{3}{2}}+\frac{x}{2\pi(1+\xi x)}\Big[\frac{d}{dx}\Big[\frac{\xi j}{1+\xi x}\\\nonumber &+&\frac{k \sqrt{\xi(2\xi^{2}+\alpha)}}{\sqrt{3\xi+(\xi^{2}-\alpha)x}\ (1+\xi x)}+\dot{M}(x)\Big]+\frac{\dot{y}^{2}}{y^{2}}\Big]
-\frac{\xi x}{4\pi(1+\xi x)^{2}}\Big[\frac{\xi j}{1+\xi x}\\ \label{5c}&+&\frac{k \sqrt{\xi(2\xi^{2}+\alpha)}}{\sqrt{3\xi+(\xi^{2}-\alpha)x}\ (1+\xi x)}+\dot{M}(x)\Big]
+\frac{\xi^{2}x}{8\pi(1+\xi x)^{2}}-\frac{\alpha x}{4\pi(1+\xi x)},
\end{eqnarray}
\begin{eqnarray}\nonumber
\Delta&=&\frac{x}{2\pi(1+\xi x)}\Big[\frac{d}{dx}\Big[\frac{\xi j}{1+\xi x}+\frac{k \sqrt{\xi(2\xi^{2}+\alpha)}}{\sqrt{3\xi+(\xi^{2}-\alpha)x}\ (1+\xi x)}+\dot{M}(x)\Big]+\frac{\dot{y}^{2}}{y^{2}}\Big]\\ \nonumber
&-&\frac{\xi x}{4\pi(1+\xi x)^{2}}\Big[\frac{\xi j}{1+\xi x}+\frac{k \sqrt{\xi(2\xi^{2}+\alpha)}}{\sqrt{3\xi+(\xi^{2}-\alpha)x}\ (1+\xi x)}+\dot{M}(x)\Big]\\ \label{6c}
&+&\frac{\xi^{2}x}{8\pi(1+\xi x)^{2}}-\frac{\alpha x}{4\pi(1+\xi x)}.
\end{eqnarray}
The function $\dot{M}(x)$ and $\frac{\dot{y}}{y}$ are defined as
\begin{eqnarray}\nonumber
\frac{\dot{y}}{y}&=&\frac{\xi j}{1+\xi x}+\frac{k \sqrt{\xi(2\xi^{2}+\alpha)}}{\sqrt{3\xi+(\xi^{2}-\alpha)x}\ (1+\xi x)}+\dot{M}(x),\\\nonumber
\dot{M}(x)&=&\frac{(\xi^{2}-\alpha)}{4\xi}(\mathcal{A}+1)+\frac{\mathcal{B}\ [(\xi^{2}-\alpha)x+3\xi]^{-\frac{1}{2}}}{16\ \xi^{2}\ \sqrt{2\pi}\ (1+\xi x)^{2}}\Big[\xi^{3}x(6\xi^{2}+2\xi^{3}x-9\alpha)+3\alpha^{2}(1\\\nonumber&+&3\xi x)
-4\xi^{4}(-3+x^{2}\alpha)+\xi^{2}\alpha(3+2x^{2}\alpha)\Big].
\end{eqnarray}
\item For $\eta=2/3$
\begin{eqnarray}\label{7c}
P_{r}&=&\mathcal{A}\ \Big[\frac{3\xi+\xi^2 x-\alpha x}{8\pi(1+\xi x)^{2}}\Big]+\mathcal{B}\ \Big[\frac{3\xi+\xi^2 x-\alpha x}{8\pi(1+\xi x)^{2}}\Big]^{\frac{5}{2}},\\\nonumber
P_{t}&=&\mathcal{A}\ \Big[\frac{3\xi+\xi^{2}x-\alpha x}{8\pi(1+\xi x)^{2}}\Big]+\mathcal{B}\ \Big[\frac{3\xi+
\xi^{2}x-\alpha x}{8\pi(1+\xi x)^{2}}\Big]^{\frac{5}{2}}+\frac{x}{2\pi(1+\xi x)}\Big[\frac{d}{dx}\Big(\frac{\xi l}{1+\xi x}\\ \nonumber&+&\frac{m \sqrt{\xi(2\xi^{2}+\alpha)}}{\sqrt{3\xi+(\xi^{2}-\alpha)x}\ (1+\xi x)}+\dot{N}(x)\Big)+\frac{\dot{y}^{2}}{y^{2}}\Big]
-\frac{\xi x}{4\pi(1+\xi x)^{2}}\Big[\frac{\xi l}{1+\xi x}\\\label{8c}&+&\frac{m \sqrt{\xi(2\xi^{2}+\alpha)}}{\sqrt{3\xi+(\xi^{2}-\alpha)x}\ (1+\xi x)}+\dot{N}(x)\Big]
+\frac{\xi^{2}x}{8\pi(1+\xi x)^{2}}-\frac{\alpha x}{4\pi(1+\xi x)},\\\nonumber
\Delta&=&\frac{x}{2\pi(1+\xi x)}\Big[\frac{d}{dx}\Big(\frac{\xi l}{1+\xi x}+\frac{m \sqrt{\xi(2\xi^{2}+\alpha)}}{\sqrt{3\xi+(\xi^{2}-\alpha)x}\ (1+\xi x)}+\dot{N}(x)\Big)+\frac{\dot{y}^{2}}{y^{2}}\Big]\\ \nonumber
&-&\frac{\xi x}{4\pi(1+\xi x)^{2}}\Big[\frac{\xi l}{1+\xi x}+\frac{m \sqrt{\xi(2\xi^{2}+\alpha)}}{\sqrt{3\xi+(\xi^{2}-\alpha)x}\ (1+\xi x)}+\dot{N}(x)\Big]\\ \label{9c}
&+&\frac{\xi^{2}x}{8\pi(1+\xi x)^{2}}-\frac{\alpha x}{4\pi(1+\xi x)}.
\end{eqnarray}
The function $\dot{N}(x)$ and $\frac{\dot{y}}{y}$ are given as
\begin{eqnarray}\nonumber
\frac{\dot{y}}{y}=\frac{\xi l}{1+\xi x}+\frac{m \sqrt{\xi(2\xi^{2}+\alpha)}}{\sqrt{3\xi+(\xi^{2}-\alpha)x}\ (1+\xi x)}+\dot{N}(x),~~~~~~~~~~~~~~~~~~~~~~~~~~~&&
\end{eqnarray}

\begin{eqnarray}\nonumber
\dot{N}(x)&=&-\frac{\mathcal{B}\ (\xi^{2}-\alpha)\ \sqrt{(3\xi+(\xi^{2}-\alpha)x)}}{1536 \sqrt{2}\ \xi^{2}\ \pi^{\frac{3}{2}}}\Big[\frac{66(\xi^{2}-\alpha)}{(1+\xi x)^{2}}+\frac{26\ (2\xi^{2}+\alpha)}{(1+\xi x)^{3}}\Big]\\\nonumber&-&\frac{\mathcal{B} (\xi^{2}-\alpha)}{3072 \sqrt{2}\ \xi^{3} \pi^{\frac{3}{2}}\sqrt{(3\xi
+(\xi^{2}-\alpha)x)}} \Big[\frac{33(\xi^{2}-\alpha)^{2}}{(1+\xi x)}+\frac{26(\xi^{2}-\alpha)(2\xi^{2}+\alpha)}{(1+\xi x)^{2}}
\\\nonumber&+&\frac{8(2\xi^{2}+\alpha)^{2}}{(1+\xi x)^{3}}\Big]+\frac{\mathcal{B}\ \sqrt{(3\xi+(\xi^{2}-\alpha)x)}}{512\ \sqrt{2}\ \xi^{2}\ \pi^{\frac{3}{2}}}\times \Big[\frac{33(\xi^{2}-\alpha)^{2}}{(1+\xi x)^{2}}
+\frac{26(\xi^{2}-\alpha)}{(1+\xi x)^{3}}(2\xi^{2}+\alpha)\\\nonumber&+&\frac{8(2\xi^{2}+\alpha)^{2}}{(1+\xi x)^{4}}\Big].
\end{eqnarray}
\item For $\eta=1/2$
\begin{eqnarray}\label{10c}
P_{r}&=&\mathcal{A}\ \left[\frac{3\xi+\xi^2 x-\alpha x}{8\pi(1+\xi x)^{2}}\right]+\mathcal{B}\ \left[\frac{3\xi+\xi^2 x-\alpha x}{8\pi(1+\xi x)^{2}}\right]^{3},\\\nonumber
P_{t}&=&\mathcal{A}\ \left[\frac{3\xi+\xi^{2}x-\alpha x}{8\pi(1+\xi x)^{2}}\right]+\mathcal{B}\ \left[\frac{3\xi
+\xi^{2}x-\alpha x}{8\pi(1+\xi x)^{2}}\right]^{3}+ \frac{x}{2\pi(1+\xi x)}\Big[\frac{n(n-1)\xi^{2}}{(1+\xi x)^{2}}\\\nonumber&+&\frac{2\ n\ \xi\ \dot{O}(x)}{(1+\xi x)}+\dot{O}(x)^{2}
+\ddot{O}(x)\Big]-\frac{\xi x}{4\pi(1+\xi x)^{2}}\left[\frac{n\ \xi}{(1+\xi x)}+\dot{O}(x)\right]
+\frac{\xi^{2} x}{8\pi(1+\xi x)^{2}}\\ \label{11c}
&-&\frac{\alpha x}{4\pi(1+\xi x)},
\end{eqnarray}
\begin{eqnarray}\nonumber
\Delta&=&\frac{x}{2\pi(1+\xi x)}\left[\frac{n(n-1)\xi^{2}}{(1+\xi x)^{2}}+
\frac{n\ \xi\ \dot{O}(x)}{(1+\xi x)}+\dot{O}(x)^{2}+\ddot{O}(x)\right]
-\frac{\xi x}{4\pi(1+\xi x)^{2}}\Big[\frac{n\ \xi}{2(1+\xi x)}\\ \label{12c}&+&\dot{O}(x)\Big]+ \frac{\xi^{2}x}{8\pi(1+\xi x)^{2}}
-\frac{\alpha x}{4\pi(1+\xi x)}.
\end{eqnarray}
The function $\dot{O}(x)$ and $\ddot{O}(x)$ are described as
\begin{equation}\nonumber
\dot{O}(x)=\frac{(\xi^{2}-\alpha)}{4\xi}\ (\mathcal{A}+1)+ \frac{\mathcal{B}}{256\pi^{2}}\left[\frac{(3\xi+\xi^{2} x-\alpha x)^{3}}{(1+\xi x)^{5}}\right],
\end{equation}
\begin{equation}\nonumber
\ddot{O}(x)=-\frac{\mathcal{B}}{256 \pi^{2}}\left[\frac{[2\xi^{2}(6+\xi x)+\alpha(3-2\xi x)](3\xi+\xi^{2} x-\alpha x)^{2}}{(1+\xi x)^{6}}\right].
\end{equation}
\end{itemize}

\end{document}